\documentclass[preprint,aps,nofootinbib,superscriptaddress,showpacs]{revtex4}
\usepackage{epsfig}
\usepackage{bm}

\def\be{\begin{equation}} \def\ee{\end{equation}} \def\bea{\begin{eqnarray}}
\def\eea{\end{eqnarray}} \def\nnb{\nonumber}

\begin{document}

\title{
Polarization of the neutron induced from hadronic weak interactions
in the photo-disintegration of the deuteron}

\author{J. W. Shin}
\affiliation{Department of Physics, Sungkyunkwan University,
Suwon 440-746, Korea}

\author{C. H. Hyun}
\email{hch@daegu.ac.kr}
\affiliation{Department of Physics Education,
Daegu University, Gyeongsan 712-714, Korea}
\affiliation{School of Physics, Korea Institute for
Advanced Study, Seoul 130-722, Korea}

\author{S.-I. Ando}
\affiliation{Department of Physics Education,
Daegu University, Gyeongsan 712-714, Korea}

\author{S.-W. Hong}
\affiliation{Department of Physics and Energy Science, Sungkyunkwan University,
Suwon 440-746, Korea}

\date{March 12, 2013}

\begin{abstract}
New observables with which we can study the
two-nucleon weak interactions at low energies are considered.
In the breakup of the deuteron by photons,
polarization of outgoing neutrons can depend on the
parity-violating component of two-nucleon interactions.
We express the parity-violating polarization in general forms,
and perform numerical calculations with a pionless effective field
theory.
The theory has unknown parity-violating low energy constants,
and the results are expressed in linear combination of them. 
We discuss the results and their implication to the understanding
of the hadronic weak interactions.

\end{abstract}

\maketitle

\section{Introduction}

Present knowledge about the most fundamental interactions
indicate that parity is not conserved only in the weak interactions.
Such a nature of the weak interaction has been successfully probed
in leptonic and semi-leptonic processes in high-energy experiments 
as well as in decay.
In principle parity-violating (PV) aspects of the weak interaction
can emerge in the pure hadronic processes too at both high and
low energies. However our understanding of PV aspects of the weak interaction
in the low energy region is 
still very poor even though more than 50 years have passed since the 
first observation of the parity violation in nuclear phenomena.

Nevertheless efforts in both experiments and theories have been continued.
Especially there has been significant progress in the low energy 
few-nucleon systems in the last three decades. In Ref.~\cite{DDH},
the authors wrote down the two-nucleon PV interactions in terms of
$\pi$-, $\rho$- and $\omega$-meson exchanges (DDH potential), which 
contain seven weak meson-nucleon coupling constants. Several PV observables
in nuclear and hadronic processes have been calculated in terms
of the DDH potential, and experiments were attempted to determine the
values of the seven weak meson-nucleon coupling constants.
In the last decade, calculations have been improved by the
use of modern nucleon-nucleon ($NN$) phenomenological potentials
such as Argonne v18, CD Bonn and Nijmegen93. 
The relevant PV observables are the asymmetry in $\vec{n} p \to d\gamma$
\cite{plb01, schi04}, 
the anapole moment of the deuteron \cite{plb03, prc03},
the polarization in $np \to d\gamma$ \cite{epja05np}, asymmetry in 
$d\vec{\gamma} \to np$ \cite{prc04, fuji04, epja05dg}, longitudinal asymmetry
in $\vec{p} p$ scattering \cite{prc06, prc08}. The PV observables were expressed
in terms of the seven meson-nucleon coupling constants in the DDH potential.
Longitudinal asymmetries were measured with good accuracies at low 
energies \cite{ever91}. 
Polarization in $np \to d\gamma$ was measured in the late 70's,
but the experiments could provide only the upper limit \cite{knya84}. 
Asymmetry in $\vec{n}p \to d\gamma$ is under measurement at SNS 
in Oak Ridge \cite{geri11}.
There was an experimental trial for the deuteron anapole moment and 
$d\vec{\gamma} \to np$.

In the mean time, there was a reformulation in the theory 
for the PV interactions in the frame work of effective field
theory (EFT) \cite{zhu05}. Authors in \cite{zhu05}  derived the $NN$ 
PV interactions from the theory with pions (pionful theory) and also
without pions (pionless theory). In the pionless theory where all
the interactions are described in terms of contact terms only,
it was shown that only five PV low energy constants (LECs) are
independent after removing redundancy in the DDH potential \cite{girl08}.
PV observables in the two-nucleon systems were recalculated with
the EFT PV potentials with and without pions 
\cite{savage01, plb07, liu07, phil09, prc10, schin10}.
Nowadays, effort to determine the PV LECs in the pionless theory
is getting more attention in the field.
Asymmetry in $d\vec{\gamma} \to np$ is now becoming a potential
candidate for the measurements in the two-nucleon processes.
Measurements have been proposed at JLab, SPring-8, Shanghai
Synchrotron, and most recently at TUNL.
The aimed accuracy in the experiment at TUNL is of the order of $10^{-8}$,
with which one can obtain stringent constraint to pin down the values
of either meson-nucleon coupling constants in the DDH potential
or the PV LECs in the pionless theory.
For precise determination of the coupling constants or LECs,
however, it is necessary to have additional observables.

In this work, we calculate polarization of the neutron in $d\gamma \to 
\vec{n} p$ at low energies. There is a long history of discrepancy
between theory and experiment for the polarization $P_{y'}$ 
\cite{rus60,sch05,kukulin08}, which is a parity-conserving (PC) quantity. 
The problem with a pionless EFT with dibaryon fields as
auxiliary fields for the two-nucleon states was revisited \cite{ando11}. 
With dibaryon fields,
the calculation becomes simple and the convergence is especially efficient
at low energies. In fact, we applied the theory successfully
to various quantities such as the electromagnetic moments of the 
deuteron \cite{ando05}, $np$ capture at the big-bang nucleosynthesis 
energies \cite{ando06b}, and $pp$ fusion in the Sun \cite{ando08}. 
Also, we observed good agreement with other theoretical results
for $P_{y'}$ at low energies, but the discrepancy with the measurements
still remains unresolved. 
$P_{y'}$ is the polarization along $y'$ axis (convention for the coordinate
system will be shown later). One can also think of the polarization along
$x'$ and $z'$ directions, but they vanish if only PC interactions 
are considered. As will be shown in the following section,
however, PV interactions cause non-zero contributions to $P_{x'}$ and 
$P_{z'}$. Motivated by this simple observation, we calculate $P_{x'}$ and
$P_{z'}$ with a pionless EFT with dibaryon fields.
Assuming the first order
approximation, the observables are obtained in the linear combination of 
PV LECs. Since the values of PV LECs are completely unknown, we cannot
determine the numerical values of the polarizations. Instead, the 
coefficients of the LECs can be calculated easily. We compare the resultant
coefficients with those appearing in other PV observables such as 
the asymmetry in $\vec{n}p \to d\gamma$ and the polarization in 
$np \to d\gamma$. By this comparison, we can roughly estimate the order 
of the physical quantity, and discuss the feasibility of the measurement.

The paper is organized as follows. In Sect.~II, we present the basic
Lagrangians. In Sect.~III, we obtain the diagrams at leading order (LO),
and calculate the amplitudes. In Sect.~IV, the 
numerical results are discussed. 
We summarize the work in Sect.~V, and give detailed forms of complicated
equations are given in Appendix A.

\section{Effective Lagrangian}

In the pionless theory, pions are treated as heavy degrees of freedom, 
and thus the typical
scale of expansion parameter is $Q/m_\pi$, where $Q$ is a physical or 
exchange momentum.
In the system where scattering length is unusually long
or binding energy is very shallow, one can also treat these small 
scales as expansion parameters.
It is natural to assign order $Q$ to the quantities such as 
$\gamma$, $1/a_0$, $1/a_1$, $1/r_0$ and $1/\rho_d$, 
where $a_0$, $a_1$ are the $np$ scattering
length in the $^1 S_0$ and $^3 S_1$ states, respectively and $r_0$ is the
effective range in the $^1 S_0$ state. $\gamma = \sqrt{m_N\, B}$ where
$B$ is the binding energy of the deuteron and $\rho_d$ is the effective
range corresponding to the deuteron.
In a diagram, propagators of a single nucleon and a dibaryon field are
counted as $Q^{-2}$ and integration of a nucleon loop generates $Q^5$.

\subsection{Parity-conserving part}

PC part of the Lagrangian consists of strong
and electromagnetic (EM) interactions.
PC Lagrangian with dibaryon fields can be written as \cite{ando05}

\begin{eqnarray}
{\cal L}_{\rm PC} = {\cal L}_N + {\cal L}_s + {\cal L}_t 
+ {\cal L}_{st},
\end{eqnarray}
where ${\cal L}_N$, ${\cal L}_s$, ${\cal L}_t$ and ${\cal L}_{st}$ 
include interactions for nucleons, dibaryon in $^1 S_0$ state, 
dibaryon in $^3 S_1$ state, and EM transition between 
$^1 S_0$ and $^3 S_1$ states, respectively. 
Retaining terms that are relevant to the present work, we have

\begin{eqnarray}
{\cal L}_N &=& 
N^\dagger \left \{ i D_{0} + \frac{{\vec{D}}^{2}}{2 m_N}
-\frac{e}{2m_{N}}\frac{1}{2}(\mu_{S}+\mu_{V}\tau_{3}) 
\vec{\sigma} \cdot \vec{B} \right \} N, 
\\
{\cal L}_s &=& 
-s^\dagger_a \left\{ i D_{0} + \frac{{\vec{D}}^{2}}{4 m_N}
+ \Delta_s \right\} s_a
- y_s \left\{ s^\dagger_a [ N^T P^{(^1 S_0)}_a N ] + {\rm h.c.} \right\}, 
\\
{\cal L}_t &=& 
-t^\dagger_i \left\{ i D_{0} + \frac{{\vec{D}}^{2}}{4 m_N}
+ \Delta_t \right\} t_i
- y_t \left\{ t^\dagger_i [ N^T P^{(^3 S_1)}_i N ] + {\rm h.c.} \right\}
-\frac{2L_{2}}{m_{N}\rho_{d}}(i) \epsilon_{ijk} t^\dagger_i t_j B_{k}, 
\\
{\cal L}_{st} &=& \frac{L_1}{m_N \sqrt{r_0 \rho_d}} 
[ t^\dagger_i s_3 B_i + {\rm h.c.} ],
\end{eqnarray}
where the projection operators for the $^1 S_0$ and $^3 S_1$ states
are respectively defined as

\begin{eqnarray}
P^{(^1 S_0)}_a = \frac{1}{\sqrt{8}} \sigma_2 \tau_2 \tau_a,  
\ \ \
P^{(^3 S_1)}_i = \frac{1}{\sqrt{8}} \sigma_2 \sigma_i \tau_2.
\end{eqnarray}
The covariant derivative is defined as
$D_\mu \equiv \partial_\mu - ieQ{\cal V}_\mu^{\rm ext}$ where
${\cal V}^{\rm ext}_\mu$ represents the external vector field.
For the nucleon, we use $D_{0} = \partial_{0}-ieQ{\cal V}_{0}^{\rm ext}$,
$\vec{D} = \vec{\nabla}+ieQ{\vec{\cal{V}}}^{\rm ext}$, 
where $Q=\frac{1}{2}(1+\tau_{3})$ is the charge operator.
For the dibaryon fields, we have 
$D_{0} = \partial_{0}-ie{\cal V}_{0}^{\rm ext}$,
$\vec{D} = \vec{\nabla}+ie{\vec{\cal{V}}}^{\rm ext}$.
Dibaryon fields in $^1 S_0$ and $^3 S_1$ states are denoted
by $s_a$ and $t_i$, respectively, and $B_i$ is the external magnetic field
given by $\vec{B} = \nabla \times \vec{{\cal V}}^{\rm ext}$.
$\Delta_{s,t}$ are defined by the mass difference between the
dibaryon and two nucleon states, i.e. 
$\Delta_{s, t} = m_{s, t} - 2 m_N$. 

LECs $y_s$ and $y_t$ represent the strength of the coupling between
a two-nucleon state and a dibaryon field. 
They are determined from the empirical values of effective range parameters,
$y_s = \frac{2}{m_N} \sqrt{\frac{2 \pi}{r_0}}$ and
$y_t = \frac{2}{m_N} \sqrt{\frac{2 \pi}{\rho_d}}$.
LECs $L_1$ and $L_2$ can be determined from the $np$ capture 
cross section at threshold and the deuteron magnetic moment,
respectively \cite{ando05}.

\subsection{Parity-violating part}

It was shown that the insertion of a nucleon loop in the 
propagator of a dibaryon field leaves the order of the diagram
the same as that of a single dibaryon propagator \cite{bean01}.
As a result, LO diagrams have only dibaryon-$NN$ ($dNN$)
vertices for the strong interaction, and other types of strong vertices,
e.g. four-nucleon contact terms belong to sub-leading contributions.
If we are to consider the weak effects, we have to include PV interactions
in a diagram. This can be easily achieved by simply replacing one PC
vertex in a diagram for PC transition with a PV interaction.
Even with this replacement, remaining part of the diagram is unchanged,
so the ordering of the diagram is not affected by the insertion of a PV vertex.
Therefore, it may suffice to represent the weak interactions in terms
of only PV $dNN$ vertices at LO.

At low energies, two-nucleon systems are dominantly occupied by
$S$-wave states, i.e., $^1 S_0$ and $^3 S_1$. 
PV interactions change the spatial parity of the $S$-wave states to the
next low lying opposite parity states such as $^3 P_J$ and $^1 P_1$.
$^1 P_1$ is isosinglet, and thus it is allowed to $np$ system only.
On the other hand, $^3 P_J$ are isotriplet, and thus $nn$ and $pp$ as well as
$np$ can occupy the states.
If we consider the change of the states from $S$-wave to $P$-wave by
the PV interaction, we have the following selections:
$^1 S_0$ to $^3 P_0$ ($nn$, $pp$, $np$), $^3 S_1$ to $^1 P_1$ ($np$),
and $^3 S_1$ to $^3 P_1$ ($np$). As a result, we have five terms for
the PV $dNN$ interactions as
\begin{eqnarray}
{\cal L}^0_{\mbox{\tiny PV}} &=&
\sum^{3}_{a=1} \frac{h^{0 s a}_d}{2 \sqrt{2\, \rho_d\, r_0}\, m_N^{5/2}} 
s^\dagger_a\, N^T 
\sigma_2 \sigma_i  \tau_2 \tau_a 
\frac{i}{2} \left(\stackrel{\leftarrow}\nabla - 
\stackrel{\rightarrow}\nabla \right)_i N +{\rm h.c.} 
\label{eq:L0s}
\\
& & + 
\frac{h^{0 t}_d}{2 \sqrt{2} \rho_d\, m_N^{5/2}} \,
t^\dagger_i\, N^T \sigma_2 \tau_2 
\frac{i}{2} \left(\stackrel{\leftarrow}\nabla - 
\stackrel{\rightarrow}\nabla \right)_i N
+{\rm h.c.},
\label{eq:L0t}
\\
{\cal L}^1_{\mbox{\tiny PV}} &=&
i \frac{h^1_d}{2 \sqrt{2} \rho_d\, m_N^{5/2}} \,
\epsilon_{ijk}\, t^\dagger_i\, N^T \sigma_2 \sigma_j \tau_2  \tau_3
\frac{i}{2} \left(\stackrel{\leftarrow}\nabla - 
\stackrel{\rightarrow}\nabla \right)_k N
+{\rm h.c.}.
\label{eq:L1}
\end{eqnarray}
Superscript in ${\cal L}_{\mbox{\tiny PV}}$ denotes the change of the
isospin accompanied in the interaction.
In Eq.~(\ref{eq:L0s}), $a =1$ and $2$ give isospin operator proportional
to $\tau_3$ and identity matrix, respectively. With the isodoublet of
the proton and the neutron, these matrices give mixture of $nn$ and $pp$
states. These terms are irrelevant in this work, and the term corresponding
to $a=3$ generates isotriplet state of the $np$ system. 
For the sake of simplicity, we disregard the constants $h^{0s1}_d$ and
$h^{0s2}_d$, and replace $h^{0s3}_d$ with $h^{0s}_d$.
Consequently we have three unknown LECs $h^{0s}_d$, $h^{0t}_d$ 
and $h^1_d$ for the coupling constants of PV interactions.

\section{Amplitude}

With the counting rules, we can arrange the pertinent Feynman diagrams
order by order. We have verified in former works that applications
to the PC processes were successful \cite{ando05, ando06b, ando08, ando11}
already at the next-to-leading order (NLO).
PC amplitude up to NLO of $d\gamma \to np$
reaction is written in the form as \cite{ando11}
\bea
A_{PC} &=&
\chi_1^\dagger \vec{\sigma}\sigma_2\tau_2\chi_2^{T\dagger}\cdot
\left\{
[\vec{\epsilon}_{(d)}\times(\hat{k}\times\vec{\epsilon}_{(\gamma)})] X_{MS}
+ \vec{\epsilon}_{(d)}\vec{\epsilon}_{(\gamma)}\cdot\hat{p}\, Y_{E}
\right\}
\nnb \\ &&
+ \chi_1^\dagger \sigma_2\tau_3\tau_2\chi_2^{T\dagger}
i\vec{\epsilon}_{(d)}\cdot (\hat{k}\times\vec{\epsilon}_{(\gamma)})\,
X_{MV}
\nnb \\ &&
+ \chi_1^\dagger \vec{\sigma}\sigma_2\tau_3\tau_2\chi_2^{T\dagger} \cdot
\left\{
\vec{\epsilon}_{(d)}\vec{\epsilon}_{(\gamma)}\cdot\hat{p}\, X_{E}
+ [\vec{\epsilon}_{(d)}\times(\hat{k}\times\vec{\epsilon}_{(\gamma)})]\,
Y_{MV}
\right\}
\nnb \\ &&
+ \chi_1^\dagger \sigma_2\tau_2\chi_2^{T\dagger} \,
i\vec{\epsilon}_{(d)}\cdot (\hat{k}\times \vec{\epsilon}_{(\gamma)})\,
Y_{MS}\,,\label{eq:pcamplitude}
\eea
where
$\vec{\epsilon}_{(d)}$ and $\vec{\epsilon}_{(\gamma)}$ are the
spin polarization vectors for the incoming deuteron and photon,
respectively, while $\chi_1^\dagger$ and $\chi_2^\dagger$ are 
the spinors of the outgoing nucleons.
$\vec{k}$ is the momentum of an incoming photon, 
$\vec{p}$ is the relative three-momentum of 
the two nucleons in the final state,
and unit vectors $\hat{k}\equiv\vec{k}/|\vec{k}|$ 
and $\hat{p}\equiv\vec{p}/|\vec{p}|$.
Details for $X$'s and $Y$'s can be found in Appendix~\ref{app:pc}.

\begin{figure}[tbp]
\epsfig{file=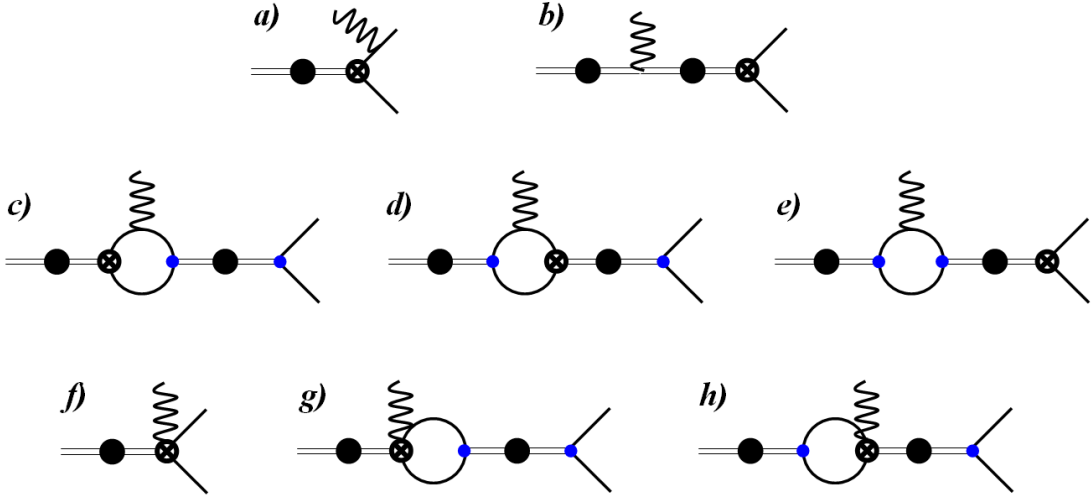, width=6.3in}
\caption{LO ($Q^{0}$) PV diagrams for $d \gamma \to \vec{n}p$.
Single solid line denotes a nucleon, a wavy line refers to a photon, and
a double line with a filled circle represents a dressed dibaryon propagator.
A circle with a cross represents a PV $dNN$ vertex.} 
\label{fig:pvdgnp}
\end{figure}

PV vertices have a spatial derivative as shown in 
Eqs.~(\ref{eq:L0s}), (\ref{eq:L0t}), (\ref{eq:L1}),
and thus they are linear in momentum. 
It is natural to count the order of a PV vertex as $Q^1$.
When a photon is coupled to a PV vertex minimal, it is equivalent to replacing the 
derivative to a photon field, and thus the order of PV minimal coupled vertices becomes $Q^0$.
With the additional counting rules for the PV vertices, the LO diagrams 
for $d\gamma \to \vec{n}p$ are obtained and depicted in Fig.~\ref{fig:pvdgnp}.
If we neglect the orders of the propagators for incoming dibaryon and outgoing nucleons, 
the diagrams are of $Q^0$.
PV amplitudes obtained from the diagrams can be written as
\begin{eqnarray}
A_{PV} = \sum_{i=a}^h A_{PV}(i).
\end{eqnarray}
Detailed expressions for $A_{PV}(i)$ are summarized in 
Appendix \ref{app:a1}.
The sum of both PC and PV contributions is
\begin{eqnarray}
A = A_{PC} + A_{PV}.
\label{eq:Atot}
\end{eqnarray}

The polarization is defined as
\begin{eqnarray}
P_{i} \equiv \frac{\sigma_{i+} - \sigma_{i-}}{\sigma_{i+} + \sigma_{i-}},
\label{eq:poldef}
\end{eqnarray}
where $\sigma_{i+}$ and $\sigma_{i-}$ are the differential cross sections
with the neutron spin up and down along a specific direction $i$, respectively.
Polarization of neutrons can be expressed by introducing the projection operator
\begin{eqnarray}
P_\pm =
\frac{1}{2} (1- \tau_3) \frac{1}{2} (1 \pm \vec{\sigma}\cdot \hat{n}),
\end{eqnarray}
where $\hat{n}$ denotes the direction of the neutron spin.
Squaring the amplitude given by Eq.~(\ref{eq:Atot}) with the polarized neutrons,
we obtain
\begin{eqnarray}
\lefteqn{
S^{-1} \sum_{spin}^{P} |A|^{2} 
= 4(|X_{MS}|^{2}+ |Y_{MV}|^{2}-2Y_{MV}{\rm Re}X_{MS}) 
}
\nonumber \\ && 
+ 2(|X_{MV}|^{2}+ |Y_{MS}|^{2}-2Y_{MS}{\rm Re}X_{MV}) 
+ 3[1-(\hat{k}\cdot \hat{p})^{2}](|X_{E}|^2+|Y_{E}|^{2}-2X_{E}Y_{E}) 
\nonumber \\ &&
\mp 2  \hat{n} \cdot (\hat{k} \times \hat{p})(X_{E}-Y_{E}){\rm Im}X_{MV} 
\mp 2 (\hat{k} \cdot \hat{n}) {\rm Im} \tilde{f}_1 
\mp 2 (\hat{p} \cdot \hat{k})(\hat{k} \cdot \hat{n}) {\rm Im} \tilde{f}_2 
\nonumber \\ && 
\mp 2 (\hat{p} \cdot \hat{n}) {\rm Im} \tilde{f}_3 
\mp 2 (\hat{p} \cdot \hat{k})(\hat{p} \cdot \hat{n}) {\rm Im} \tilde{f}_4,
\label{eq:square}
\end{eqnarray}
where $S$ is a symmetry factor for spin average, $S=2$, and $\tilde{f}_i$'s
are the PV-PC interference terms, whose details can be found in Appendix~\ref{appendixB}.

Conventions for the coordinate systems are quoted from \cite{rus60}.
We have the incoming photons along $\hat{k} = (0,0,1)$,
relative momentum of the nucleons along 
$\hat{p} = (\sin{\theta}\cos{\phi},\sin{\theta}\sin{\phi},\cos{\theta})$,
and orthogonal basis vectors are definded as
$\hat{x}' = (\cos \theta \cos \phi,\, \cos \theta \sin \phi,\, -\sin\theta)$,
$\hat{y}' = (-\sin\phi,\, \cos\phi,\, 0)$ and 
$\hat{z}' = (\sin{\theta}\cos{\phi},\sin{\theta}\sin{\phi},\cos{\theta})$.
If we align the neutron spin along $\hat{y}'$, scalar products of the unit vectors for
the PV-PC interference terms $\tilde{f}_i$ vanish. In this case, we obtain the PC 
polarization $P_{y'}$ \cite{ando11}.
If we polarize the neutrons along $\hat{n} = \hat{x}'$, 
$\hat{k}\cdot\hat{x}' = - \sin\theta$ while $\hat{p} \cdot\hat{x}'=0$,
and thus $\tilde{f}_1$ and $\tilde{f}_2$ terms are non-vanishing.
With $\hat{n} = \hat{z}'$, all the $\tilde{f}_i$ terms contribute to the
polarization $P_{z'}$. With $\hat{n} = \hat{x}'$ and $\hat{z}'$,
one can easily check that PC interference term proportional to
$\hat{n}\cdot(\hat{k}\times\hat{p})$ becomes null.
Consequently, we can obtain hadronic weak effects by calculating
the polarizations $P_{x'}$ and $P_{z'}$ which are not interfered by
the PC components of the interactions.

\section{Result and discussion}
\subsection{Polarization along $\hat{z}'$}

In this section, we present and discuss the results for $P_{z'}$.
With Eqs.~(\ref{eq:poldef}), (\ref{eq:square}), we obtain $P_{z'}$ as
\begin{eqnarray}
P_{z'} = (-2) {\rm Im}[ (\tilde{f}_1+ \tilde{f}_4 ) \cos{\theta}
+ \tilde{f}_2 \cos^{2}{\theta}+ \tilde{f}_3] / \Sigma_{PC},
\end{eqnarray}
where 
\begin{eqnarray}
\Sigma_{PC} &\equiv&  4(|X_{MS}|^{2}+ |Y_{MV}|^{2}-2Y_{MV}{\rm Re}X_{MS})
+ 2(|X_{MV}|^{2}+ |Y_{MS}|^{2}-2Y_{MS}{\rm Re}X_{MV}) 
\nonumber \\
& & + 3(1-\cos^{2}{\theta})(|X_{E}|^2+|Y_{E}|^{2}-2X_{E}Y_{E}) .
\end{eqnarray}
Since $\tilde{f}_i$'s contain linear combinations of $h^T_d$'s
($T= 0t,\,  0s,\, 1$)
whose values are not known, we may rewrite the polarization in the form
\begin{eqnarray}
P_{z'} \equiv c^{0t}_{z} h^{0t}_d + c^{0s}_{z} h^{0s}_d + c^{1}_{z} h^1_d.
\end{eqnarray}
Coefficients $c^T_z$ 
are functions of the colatitude angle $\theta$ and the relative
momentum $p$ (or equivalently photon energy in the lab frame
$E^{lab}_\gamma$),
and they take into account the characteristics of PV as well as PC interactions
of the theory. Explicit forms of $c^T_z$ can be found in Appendix~\ref{AppendixC}.
\begin{figure}[tbp]
\begin{center}
\epsfig{file=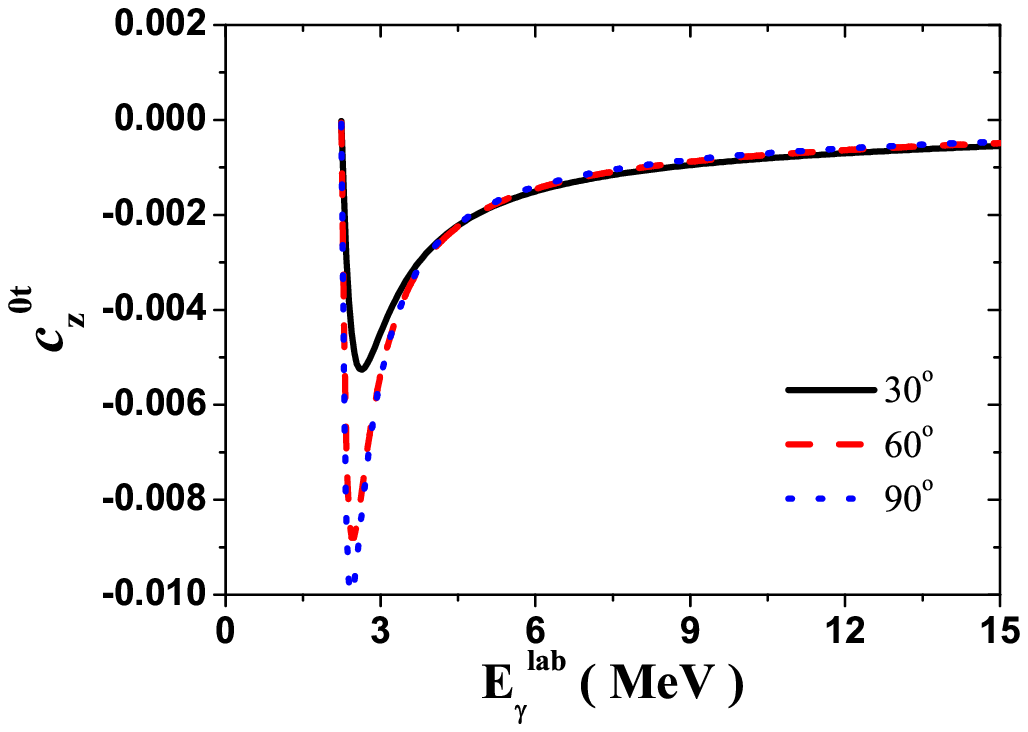, width=3.3in}
\epsfig{file=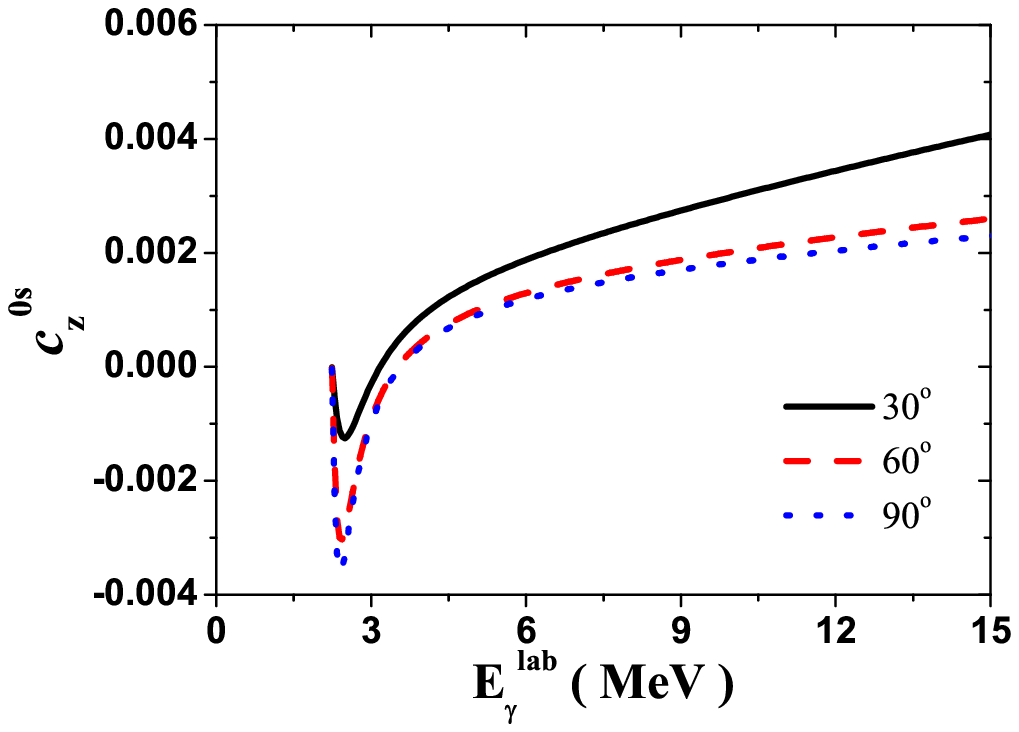, width=3.3in}
\epsfig{file=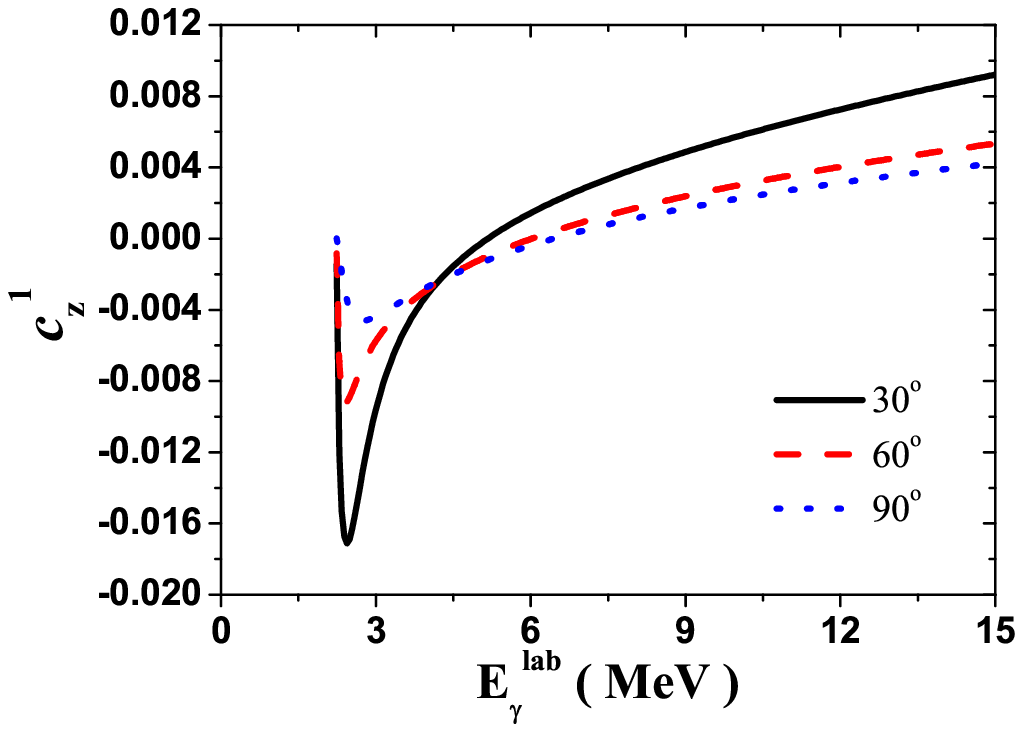, width=3.3in}
\end{center}
\caption{From the top $c^{0t}_{z}$, $c^{0s}_{z}$ and $c^{1}_{z}$ 
as functions of $E^{lab}_{\gamma}$ at $\theta_{lab}=30^\circ$, 
60$^\circ$, and 90$^\circ$.} 
\label{fig:pvz1}
\end{figure}
\begin{figure}[tbp]
\epsfig{file=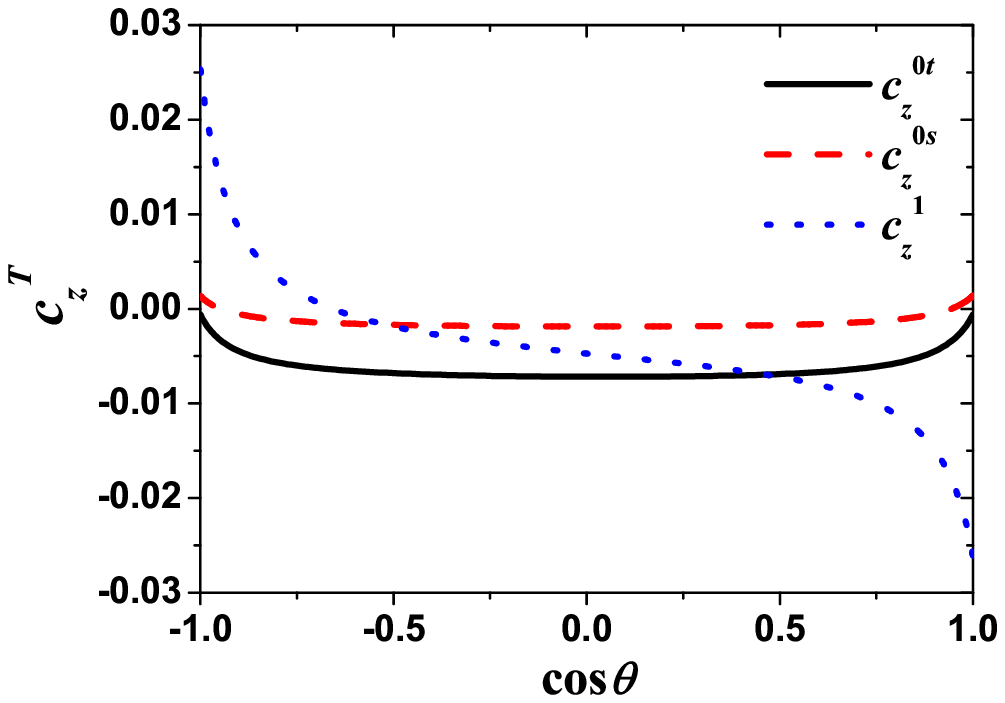, width=3.3in}
\caption{$c^{0t}_z$, $c^{0s}_z$ and $c^1_z$ as functions of 
$\cos \theta$ for $E^{lab}_{\gamma} = $ 2.75~MeV.} 
\label{fig:pvc1}
\end{figure}

In Fig.~\ref{fig:pvz1}, we plot the numerical results for 
$c^T_z$'s as functions of photon energies in the lab frame.
Angle dependences are examined by picking up three angles,
$\theta_{lab} = 30^\circ,\, 60^\circ,$ and $90^\circ$.
A common feature in $c^T_z$ is that there is a minimum in the range
$E^{lab}_\gamma = 2.4\sim2.8$~MeV regardless of angle and
the isospin structure of the PV vertex, i.e. the superscript $T$ in $h^T_d$.
At higher energies, $c^{0t}_z$ tends to converge to a value,
while $c^{0s}_z$ and $c^1_z$ show a linear increase.
Another noticeable behavior at high energies is that $c^{0s}_z$ and $c^1_z$ show
distinct dependence on the angle, but $c^{0t}_z$ is almost independent 
of the angle, 
and the magnitude of the coefficients $c^{0t}_z$ and $c^1_z$ is greater than that of
$c^{0s}_z$ by an order.

In Fig.~\ref{fig:pvc1}, we investigate in more detail 
the dependence on the angle $\theta$ in the center of mass frame
at $E^{lab}_\gamma = 2.75$~MeV.
It was seen in Fig.~\ref{fig:pvz1} that $c^T_z$'s have the largest magnitude 
in $E^{lab}_\gamma = 2.4\sim2.8$~MeV, and thus the value 2.75~MeV was chosen
arbitrarily.
$c^{0s}_z$ is almost zero regardless of the angle.
Indeed, the order of magnitude of $c^{0s}_z$ and $c^{0t}_z$ does not exceed
$7.5\times 10^{-3}$ in the energy range considered, and thus the dominance of
$c^1_z$ is very clear at forward or backward angles,
where the magnitude is greater than $2\times 10^{-2}$.
Since the values of $h^T_d$ are not known, we cannot determine
the value of $P_{z'}$.
However, we can roughly estimate the order of magnitude of $P_{z'}$
in comparison with other PV observables.

In Ref.~\cite{savage01}, $A_\gamma$ in $\vec{n}p \to d\gamma$ was calculated
with the dibaryon fields and the result was obtained as
\begin{eqnarray}
A_\gamma &=& - \frac{m^{3/2}_N}{2 \sqrt{2\pi}} h^{(1)}_{33}
\frac{1-\gamma\, a_1/3}{\kappa_1 (1-\gamma\, a_0) - \gamma^2\, a_0\, L_1/2},
\end{eqnarray}
where $h^{(1)}_{33}$ is the convention for the PV $dNN$ LEC in the work.
$h^{(1)}_{33}$ and $h^1_d$ are related through
\begin{eqnarray}
h^{(1)}_{33} = \frac{h^1_d}{\rho^{1/2}_d m^2_N},
\end{eqnarray}
which allows us to write $A_\gamma$ in terms of $h^1_d$ as
\begin{eqnarray}
A_\gamma = -3.2\times10^{-3} h^1_d.
\end{eqnarray}
The measured value of $A_\gamma$ is $-(1.5\pm 4.8)\times10^{-8}$ \cite{alberi88},
and NPDGamma collaboration aims at determining the value unambiguously at
the order of $10^{-8}$.
Since $P_{z'}$ can be larger than $A_\gamma$ by an order of magnitude, it can
be an observable experimentally advantageous in measurement to determine 
PV LEC $h^1_d$ with a better accuracy than one can achieve
with $A_\gamma$.

\subsection{Polarization along $\hat{x}'$}

Polarization along the $x'$ axis is obtained as
\begin{eqnarray}
P_{x'} =
2 \sin{\theta}{\rm Im} [ \tilde{f}_1 + \tilde{f}_2 \cos{\theta} ]  /\Sigma_{PC}.
\end{eqnarray}
We can cast it in the form
\begin{eqnarray}
P_{x'} \equiv c^{0t}_{x} h^{0t}_d + c^{0s}_{x} h^{0s}_d + c^{1}_{x} h^1_d,
\end{eqnarray}
and investigate $c^T_x$'s. Complete expressions for $c^T_x$ are shown
in Appendix~\ref{AppendixC}.

Figure~\ref{fig:pvx1} shows the energy dependence of $c^T_x$'s at the angles
$\theta_{lab} = 30^\circ,\, 60^\circ$, and $90^\circ$.
$c^{0t}_x$ and $c^{0s}_x$ show the behavior and order of magnitudes similar to those 
of $c^{0t}_z$ and $c^{0s}_z$, respectively.
The magnitude of $c^1_x$ is smaller than $c^1_z$ by a factor of $0.5\sim0.6$,
and shows a structure with a maximum and a minimum slightly
below and above 3~MeV, respectively.
Similar to $c^T_z$'s, $c^T_x$'s have a maximum magnitude for
$E^{lab}_\gamma= 2.3\sim2.7$~MeV in the low energy region.
\begin{figure}[t]
\begin{center}
\epsfig{file=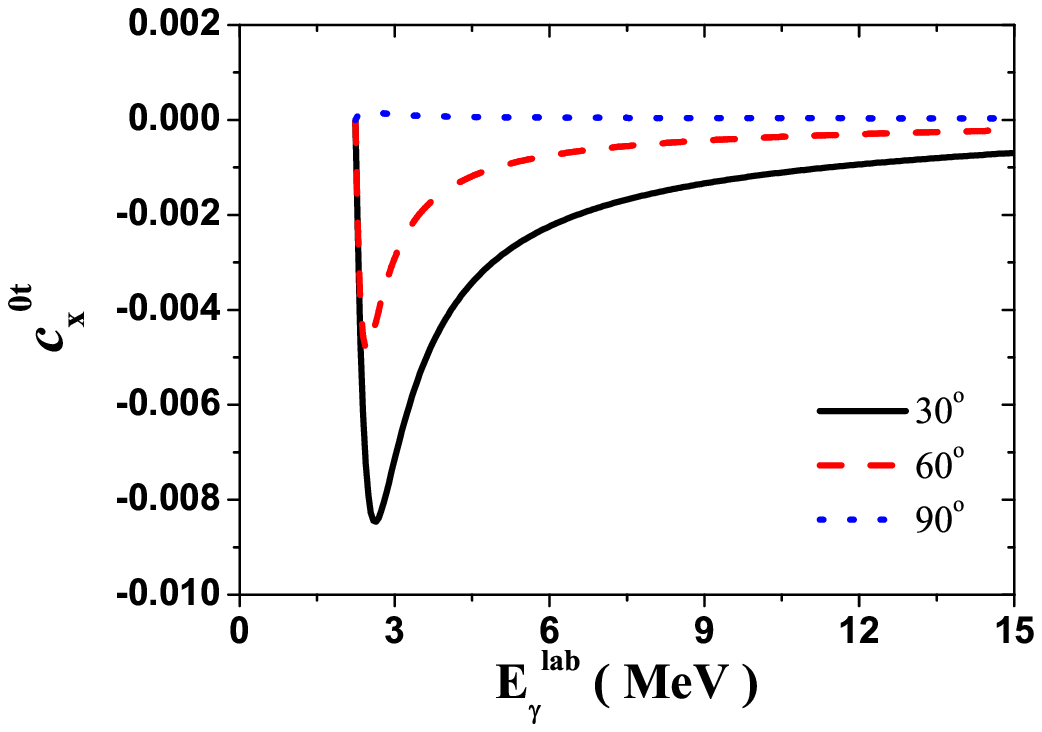, width=3.3in}
\epsfig{file=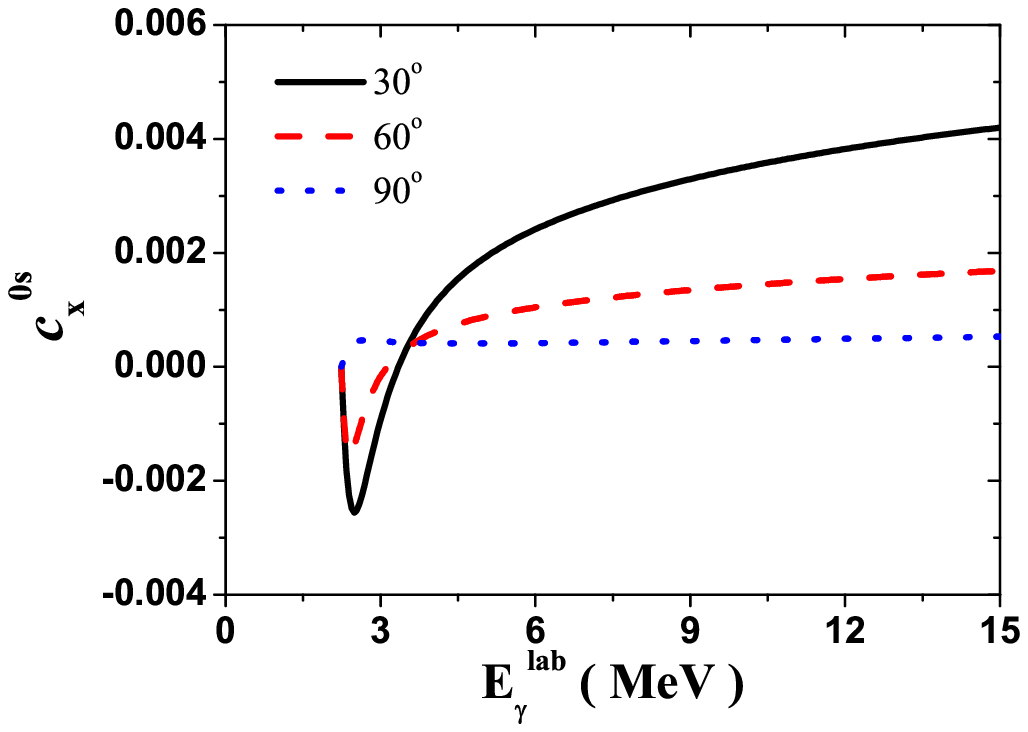, width=3.3in}
\epsfig{file=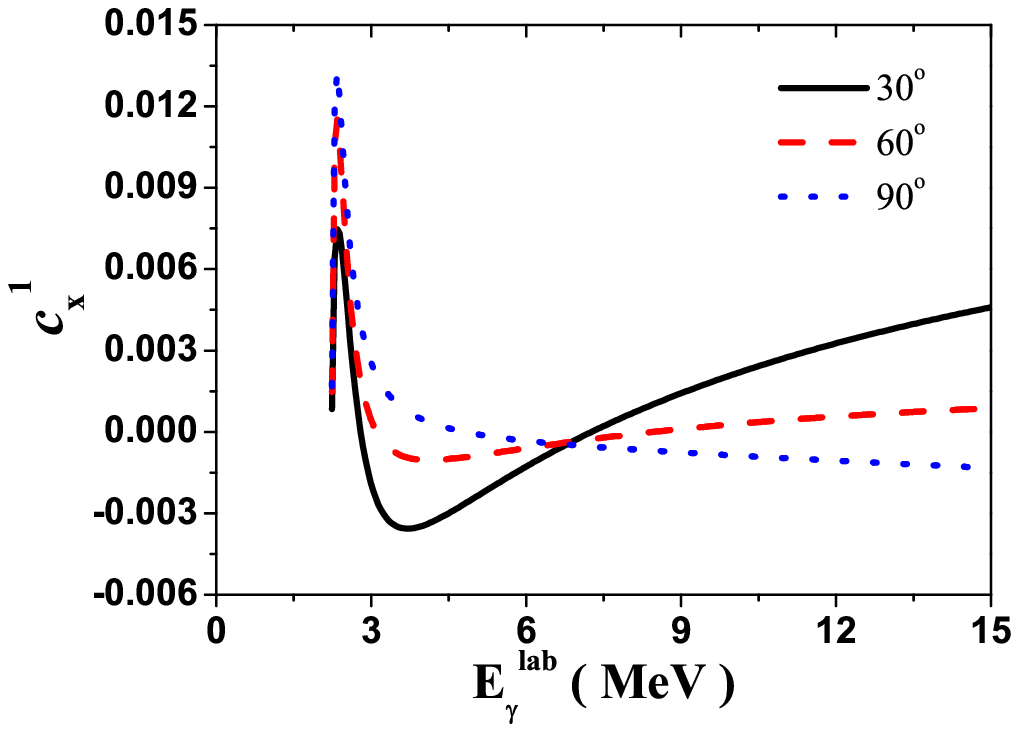, width=3.3in}
\end{center}
\caption{From the top $c^{0t}_{x}$, $c^{0s}_{x}$ and $c^{1}_{x}$ 
as functions of $E^{lab}_{\gamma}$ at $\theta_{lab}=30^\circ$, 
60$^\circ$, 90$^\circ$.} 
\label{fig:pvx1}
\end{figure}

\begin{figure}
\epsfig{file=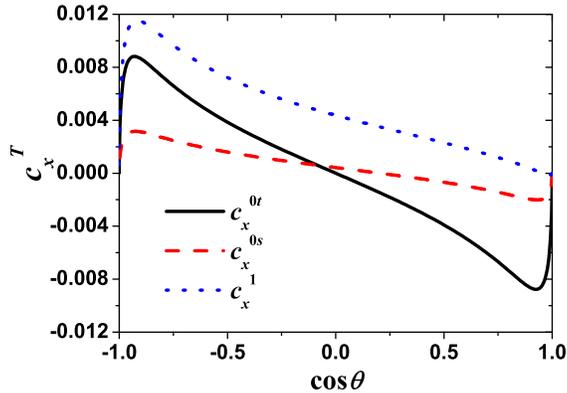, width=3.3in}
\caption{$c^{0t}_x$, $c^{0s}_x$, and $c^1_x$ as functions of 
$\cos \theta$ for $E^{lab}_{\gamma} = $ 2.75~MeV.} 
\label{fig:pvc2}
\end{figure}

In Fig.~\ref{fig:pvc2}, we plot $c^T_x$ with respect to $\cos\theta$
at $E^{lab}_\gamma=2.75$~MeV.
While $c^1_z$ monotonically dominates over other $c^T_z$ contributions,
$c^T_x$'s contain more information.
$c^{0t}_x$ forms a maximum at $\theta \simeq 22^\circ$ with the value
$c^{0t}_x \simeq -8.8\times10^{-3}$.
The values of $c^{0s}_x$ and $c^1_x$ at this angle are $-2.0\times 10^{-3}$
and $2.0\times 10^{-4}$, respectively, and thus the contribution from
$h^1_d$ can be safely ruled out in the consideration.
In Ref.~\cite{prc10}, we calculated PV polarization $P_\gamma$ in $np \to d\gamma$ at
threshold with the theory as employed in this work, 
and obtained the result  
\begin{eqnarray}
P_\gamma = - (2.59 h^{0t}_d - 1.01 h^{0s}_d) \times 10^{-2}.
\end{eqnarray}
The coefficient of $h^{0t}_d$ in $P_\gamma$ is larger than $c^{0t}_x$ in
$P_{x'}$ at $\theta \simeq 22^\circ$ by a factor of 3, 
but they are roughly
similar in order.
Therefore, measurement of $P_{x'}$ at the $\theta \simeq 22^\circ$,
in addition to PV $P_\gamma$ in $np \to d\gamma$, can provide a
complementary constraints to determine $h^{0t}_d$ and $h^{0s}_d$.
At $\theta = 90^\circ$,
$c^{0t}_x$ and $c^{0s}_x$ are of the order of $10^{-4}$ or less, while
$c^1_x \simeq 4.4\times 10^{-3}$. At this angle, we can neglect the
contributions from $c^{0t}_x$ and $c^{0s}_x$, and thus have chances
to determine $h^1_d$ with a minor uncertainty.
This value is comparable to $A_\gamma \simeq -3.3\times10^{-3}h^1_d$
in $\vec{n}p \to d\gamma$.
With the measurements of $A_\gamma$ in $\vec{n}p\to d\gamma$,
$P_{z'}$ at forward or backward angles, and $P_{x'}$ at around
$\theta = 90^\circ$, we can check the consistency of the theory
and determine the value of $h^1_d$.
At the angles around $\theta=157^\circ$, all the $c^T_x$'s become
maximum with $c^1_x\simeq 1.15\times 10^{-2}$,
and the ratios of $c^{0t}_x$ and $c^{0s}_x$ to $c^1_x$ are
0.76 and 0.27, respectively. 
By determining $h^{0t}_d$ from $P_{x'}$ at $\theta\simeq22^\circ$,
and $h^1_d$ from $A_\gamma$, $P_{z'}$ at $\theta \simeq 0^\circ$
or $180^\circ$, and $P_{x'}$ at $\theta \simeq 90^\circ$,
we may pin down the value of $h^{0s}_d$ through the measurement of 
$P_{x'}$ at the angles in the backward direction.

\section{Summary}

In this work, we have considered the polarization of the neutron in 
$d\gamma \to \vec{n}p$ with a pionless EFT incorporating 
dibaryon fields.
Polarization along the azimuthal direction $y'$ imposes the information
about the interactions that conserve parity.
Along the radial ($z'$) and colatitude ($x'$) directions,
on the other hand,
non-vanishing contributions reflect the effect of PV interactions.
We focused on the PV components of the polarization, and calculated
$P_{x'}$ and $P_{z'}$ as functions of the incident photon energies up
to 15 MeV.
Since the coefficients $c^{T}_z$'s and $c^{T}_x$'s can be evaluated at
different angles and energies, one can determine the unknown PV 
LECs by comparing the calculated $P_{z'}$ and $P_{x'}$ 
with the experimental values.

In the dependence on the energy, both $P_{x'}$ and $P_{z'}$
show a peak structure slightly above the threshold regardless
of the angle $\theta$.
At these energies, pionless EFT with dibaryon fields was verified
to a good accuracy for many observables in the two-nucleon processes,
but care should be taken as we increase energy near $\sim$ 15~MeV \cite{prc12}.
Therefore, if measurement is performed, it may be most 
desirable to concentrate on the low energies.

By fixing the photon energy to 2.75~MeV, we explored the dependence 
of $c^{T}_z$ and $c^{T}_x$ on the angle $\theta$.
Concerning $P_{z'}$, the coefficients of the iso-scalar components of the 
PV interaction, $c^{0s}_z$ and $c^{0t}_z$, are more or less constant in 
the angle, but the coefficient of the iso-vector component, $c^{1}_z$, 
changes drastically in the forward and backward angles.
In these directions, $P_{z'}$ is exclusively dominated by the
PV iso-vector interaction, and thus the measurement of $P_{z'}$ along with
$A_\gamma$ in $\vec{n}p \to d\gamma$ will provide a chance
for a unique determination of $h^1_d$.
$P_{x'}$ is expected to give more information about the PV LECs.
At the angles close to the forward direction, 
the values of $c^{0t}_x$ are more significant than those of $c^{0s}_x$ and $c^{1}_x$,
and thus $P_{x'}$ is expected to be dominated by $h^{0t}_d$.
On the other hand, in the backward
directions, contributions from $h^1_d$, $h^{0t}_d$ and $h^{0s}_d$ terms are expected 
to be of the same order.
By combining the measurements of $P_{x'}$ and $P_{z'}$ at various angles,
we can determine the PV LECs in the $np$ system,
$h^{0t}_d$, $h^{0s}_d$ and $h^1_d$.

Enhancement of $P_{z'}$ compared to $A_\gamma$ in $\vec{n}p \to d\gamma$
by an order of magnitude is a striking result.
Since the prediction can imply significant impact on the experiments, 
the results in this work should be counter-checked by other calculations.
One possibility is to adopt the DDH potential,
and obtain the results for $P_{z'}$ and $P_{x'}$ in terms of the 
PV meson-nucleon coupling constants.
Many PV observables in the two-nucleon systems were already calculated 
in terms of the DDH potential.
The work is in progress  for evaluating $P_{z'}$ and $P_{x'}$ \cite{liu13}.
%

\section*{Acknowledgments}
Work of CHH and SIA is supported by the Basic Science Research 
Program through the National Research Foundation of Korea (NRF)
funded by the Ministry of Education, Science and Technology 
Grant No. 2010-0023661, the work of JWS by 
Basic Science Research Program through the National Research 
Foundation of Korea (NRF) funded by the Ministry of Education, 
Science and Technology Grant No. 2011-0025116
and the work of SWH by 
Basic Science Research Program through the National Research 
Foundation of Korea (NRF) funded by the Ministry of Education, 
Science and Technology Grant No. 2012R1A1A2007826.

\appendix
\section{Summary of lengthy expressions}
\label{appendixA}

\subsection{PC terms}
\label{app:pc}
For the PC amplitude given in Eq.~(\ref{eq:pcamplitude}),
we have the PC terms
\bea
X_{MV} &=& - \sqrt{\frac{\pi\gamma}{1-\gamma\rho_d}}
\frac{1}{
\frac{1}{a_0}
+ip
-\frac12r_0p^2}
\frac{1}{2m_N}
\nnb \\ && \times
\left\{
\mu_V\left[
{\rm arccos}\left(
\frac{m_N}{\sqrt{
(m_N+\frac12\omega_\gamma)^2-p^2
}}
\right)
+i\ln\left(
\frac{m_N+\frac12\omega_\gamma +p}{\sqrt{
(m_N+\frac12\omega_\gamma)^2-p^2
}}
\right)
\right]
\right.
\nnb \\ && \left.
- \frac{\mu_V}{m_N} \left(
\frac{1}{a_0}
+ ip
-\frac12r_0 p^2
\right) F^+
+ \omega_\gamma L_1
\right\}\,,
\\
X_{MS} &=& - \sqrt{\frac{\pi\gamma}{1-\gamma\rho_d}}
\frac{1}{
\gamma
+ip
-\frac12\rho_d(\gamma^2+p^2)}
\frac{1}{2m_N}
\nnb \\ && \times
\left\{
\mu_S\left[
{\rm arccos}\left(
\frac{m_N}{\sqrt{
(m_N+\frac12\omega_\gamma)^2-p^2
}}
\right)
+i\ln\left(
\frac{m_N+\frac12\omega_\gamma +p}{\sqrt{
(m_N+\frac12\omega_\gamma)^2-p^2
}}
\right)
\right]
\right.
\nnb \\ && \left.
- \frac{\mu_S}{m_N} \left[
\gamma 
+ ip
-\frac12\rho_d(\gamma^2+p^2)
\right] F^+
+ 2 \omega_\gamma L_2
\right\}\,,
\\
X_{E} &=& \sqrt{
\frac{\pi\gamma}{1-\gamma\rho_d}
} \frac{1}{m_N^2}
\frac{p}{\omega_\gamma}F^+\,,
\ \ \
Y_{E} = \sqrt{
\frac{\pi\gamma}{1-\gamma\rho_d}
} \frac{1}{m_N^2}
\frac{p}{\omega_\gamma}F^-\,,
\\
Y_{MV} &=& \sqrt{
\frac{\pi\gamma}{1-\gamma\rho_d}
} \frac{\mu_V}{2m_N^2} F^-\,,
\ \ \
Y_{MS} = \sqrt{
\frac{\pi\gamma}{1-\gamma\rho_d}
} \frac{\mu_S}{2m_N^2} F^- ,
\eea
where
$\omega_\gamma$ is the incident photon energy in the c.m. frame, and
\begin{eqnarray}
F^{\pm} &=& \frac{1}{2}\left[ 
\frac{1}{1+\frac{\omega_{\gamma}}{2m_{N}}-\frac{\vec{p}\cdot \hat{k}}{m_{N}}}
\pm \frac{1}{1+\frac{\omega_{\gamma}}{2m_{N}}+\frac{\vec{p}\cdot \hat{k}}{m_{N}}} \right].
\end{eqnarray}

\subsection{PV amplitudes}
\label{app:a1}

We calculated the amplitude for the diagrams (a) to (h) in 
Fig.~\ref{fig:pvdgnp}.
${\chi}^{\dagger}_{1}$ and ${\chi}^{T \dagger}_{2}$ are the 
spinors of the nucleons in the final state.
\begin{eqnarray}
iA_{PV}(a) &=& 
C h^{0t}_d \bigg{[} 
(i){\chi}^{\dagger}_{1}\sigma_{2}\tau_{2}{\chi}^{T \dagger}_{2}
{\epsilon_{(d)}}_{i}{\epsilon_{(\gamma)}}_{j}
{\hat{k}}_{i}{\hat{p}}_{j}
\frac{p}{m_{N}}F^{+} 
\nonumber \\ & &
- (i){\chi}^{\dagger}_{1}\sigma_{2}\tau_{2}{\chi}^{T \dagger}_{2}
{\epsilon_{(d)}}_{i}{\epsilon_{(\gamma)}}_{j}
{\hat{p}}_{i}{\hat{p}}_{j}
\frac{2p^{2}}{m_{N}\omega_{\gamma}}F^{-} 
\nonumber \\
& &+ (i){\chi}^{\dagger}_{1}\sigma_{2}\tau_{3}\tau_{2}{\chi}^{T \dagger}_{2}
{\epsilon_{(d)}}_{i}{\epsilon_{(\gamma)}}_{j}
{\hat{k}}_{i}{\hat{p}}_{j}
\frac{p}{m_{N}}F^{-} 
\nonumber \\ & &
- (i){\chi}^{\dagger}_{1}\sigma_{2}\tau_{3}\tau_{2}{\chi}^{T \dagger}_{2}
{\epsilon_{(d)}}_{i}{\epsilon_{(\gamma)}}_{j}
{\hat{p}}_{i}{\hat{p}}_{j}
\frac{2p^{2}}{m_{N}\omega_{\gamma}}F^{+} 
\nonumber \\
& &+ {\chi}^{\dagger}_{1}\sigma_{i}\sigma_{2}\tau_{2}{\chi}^{T \dagger}_{2}
\epsilon_{ijk}{\epsilon_{(d)}}_{a}{\epsilon_{(\gamma)}}_{k}
{\hat{k}}_{j}{\hat{p}}_{a}
\frac{\mu_{S}}{m_{N}}pF^{-} 
\nonumber \\ & &
- {\chi}^{\dagger}_{1}\sigma_{i}\sigma_{2}\tau_{2}{\chi}^{T \dagger}_{2}
\epsilon_{ijk}{\epsilon_{(d)}}_{a}{\epsilon_{(\gamma)}}_{k}
{\hat{k}}_{j}{\hat{k}}_{a}
\frac{\mu_{S}}{2m_{N}}\omega_{\gamma}F^{+} 
\nonumber \\
& &+ {\chi}^{\dagger}_{1}\sigma_{i}\sigma_{2}\tau_{3}\tau_{2}{\chi}^{T \dagger}_{2}
\epsilon_{ijk}{\epsilon_{(d)}}_{a}{\epsilon_{(\gamma)}}_{k}
{\hat{k}}_{j}{\hat{p}}_{a}
\frac{\mu_{V}}{m_{N}}pF^{+}
\nonumber \\ & & 
- {\chi}^{\dagger}_{1}\sigma_{i}\sigma_{2}\tau_{3}\tau_{2}{\chi}^{T \dagger}_{2}
\epsilon_{ijk}{\epsilon_{(d)}}_{a}{\epsilon_{(\gamma)}}_{k}
{\hat{k}}_{j}{\hat{k}}_{a}
\frac{\mu_{V}}{2m_{N}}\omega_{\gamma}F^{-}  \bigg{]}  
\nonumber \\
&+& C h^{1}_d \bigg{[}
{\chi}^{\dagger}_{1}\sigma_{i}\sigma_{2}\tau_{2}{\chi}^{T \dagger}_{2}
\epsilon_{ijk}{\epsilon_{(d)}}_{j}{\epsilon_{(\gamma)}}_{a}
{\hat{p}}_{k}{\hat{p}}_{a}
\frac{2p^{2}}{m_{N}\omega_{\gamma}}F^{+}  
\nonumber \\ & &
- {\chi}^{\dagger}_{1}\sigma_{i}\sigma_{2}\tau_{2}{\chi}^{T \dagger}_{2}
\epsilon_{ijk}{\epsilon_{(d)}}_{j}{\epsilon_{(\gamma)}}_{a}
{\hat{k}}_{k}{\hat{p}}_{a}
\frac{p}{m_{N}}F^{-}  
\nonumber \\
&&
+ {\chi}^{\dagger}_{1}\sigma_{i}\sigma_{2}\tau_{3}\tau_{2}{\chi}^{T \dagger}_{2}
\epsilon_{ijk}{\epsilon_{(d)}}_{j}{\epsilon_{(\gamma)}}_{a}
{\hat{p}}_{k}{\hat{p}}_{a}
\frac{2p^{2}}{m_{N}\omega_{\gamma}}F^{-}  
\nonumber \\ & &
- {\chi}^{\dagger}_{1}\sigma_{i}\sigma_{2}\tau_{3}\tau_{2}{\chi}^{T \dagger}_{2}
\epsilon_{ijk}{\epsilon_{(d)}}_{j}{\epsilon_{(\gamma)}}_{a}
{\hat{k}}_{k}{\hat{p}}_{a}
\frac{p}{m_{N}}F^{+}  
\nonumber \\
&&+ (i){\chi}^{\dagger}_{1}\sigma_{2}\tau_{2}{\chi}^{T \dagger}_{2}
{\epsilon_{(d)}}_{i}{\epsilon_{(\gamma)}}_{i}
\left(\frac{\mu_{V}}{2m_{N}}\omega_{\gamma}F^{-}
- {\hat{p}}_{j}{\hat{k}}_{j}\frac{\mu_{V}}{m_{N}}pF^{-}\right) 
\nonumber \\ & &
+ (i){\chi}^{\dagger}_{1}\sigma_{2}\tau_{2}{\chi}^{T \dagger}_{2}
{\epsilon_{(d)}}_{i}{\epsilon_{(\gamma)}}_{j}
{\hat{k}}_{i}{\hat{p}}_{j}
\frac{\mu_{V}}{m_{N}}pF^{+} 
\nonumber \\
& &+ {\chi}^{\dagger}_{1}\sigma_{i}\sigma_{2}\tau_{2}{\chi}^{T \dagger}_{2}
\epsilon_{abc}{\epsilon_{(d)}}_{c}{\epsilon_{(\gamma)}}_{i}
{\hat{k}}_{a}{\hat{p}}_{b}
\frac{\mu_{V}}{m_{N}}pF^{-} 
\nonumber \\ & &
- {\chi}^{\dagger}_{1}\sigma_{i}\sigma_{2}\tau_{2}{\chi}^{T \dagger}_{2}
\epsilon_{abc}{\epsilon_{(d)}}_{c}{\epsilon_{(\gamma)}}_{a}
{\hat{k}}_{i}{\hat{p}}_{b}
\frac{\mu_{V}}{m_{N}}pF^{-} 
\nonumber \\
& &+ {\chi}^{\dagger}_{1}\sigma_{i}\sigma_{2}\tau_{2}{\chi}^{T \dagger}_{2}
\epsilon_{abc}{\epsilon_{(d)}}_{c}{\epsilon_{(\gamma)}}_{a}
{\hat{k}}_{i}{\hat{k}}_{b}
\frac{\mu_{V}}{2m_{N}}\omega_{\gamma}F^{+} 
\nonumber \\
& &+ (i){\chi}^{\dagger}_{1}\sigma_{2}\tau_{3}\tau_{2}{\chi}^{T \dagger}_{2}
{\epsilon_{(d)}}_{i}{\epsilon_{(\gamma)}}_{i}
\left(\frac{\mu_{S}}{2m_{N}}\omega_{\gamma}F^{+}
- {\hat{p}}_{j}{\hat{k}}_{j}\frac{\mu_{S}}{m_{N}}pF^{-}\right) 
\nonumber \\ & &
+ (i){\chi}^{\dagger}_{1}\sigma_{2}\tau_{3}\tau_{2}{\chi}^{T \dagger}_{2}
{\epsilon_{(d)}}_{i}{\epsilon_{(\gamma)}}_{j}
{\hat{k}}_{i}{\hat{p}}_{j}
\frac{\mu_{S}}{m_{N}}pF^{-} 
\nonumber \\ &&
+ {\chi}^{\dagger}_{1}\sigma_{i}\sigma_{2}\tau_{3}\tau_{2}{\chi}^{T \dagger}_{2}
\epsilon_{abc}{\epsilon_{(d)}}_{c}{\epsilon_{(\gamma)}}_{i}
{\hat{k}}_{a}{\hat{p}}_{b}
\frac{\mu_{S}}{m_{N}}pF^{+} 
\nonumber \\ & &
- {\chi}^{\dagger}_{1}\sigma_{i}\sigma_{2}\tau_{3}\tau_{2}{\chi}^{T \dagger}_{2}
\epsilon_{abc}{\epsilon_{(d)}}_{c}{\epsilon_{(\gamma)}}_{a}
{\hat{k}}_{i}{\hat{p}}_{b}
\frac{\mu_{S}}{m_{N}}pF^{+} 
\nonumber \\ &&
+ {\chi}^{\dagger}_{1}\sigma_{i}\sigma_{2}\tau_{3}\tau_{2}{\chi}^{T \dagger}_{2}
\epsilon_{abc}{\epsilon_{(d)}}_{c}{\epsilon_{(\gamma)}}_{a}
{\hat{k}}_{i}{\hat{k}}_{b}
\frac{\mu_{S}}{2m_{N}}\omega_{\gamma}F^{-} \bigg{]} \,,
\\
iA_{PV}(b) &=& 
C p\, \omega_{\gamma}\bigg[
h^{0s}_d\, d'_{s}L_{1}{\chi}^{\dagger}_{1}\sigma_{i}\sigma_{2}
\tau_{3}\tau_{2}{\chi}^{T \dagger}_{2}
\epsilon_{abc}{\epsilon_{(d)}}_{a}{\epsilon_{(\gamma)}}_{b}
\hat{k}_{c}\hat{p}_{i} 
\nonumber \\ && 
+ h^{0t}_d\, d'_{t}(i)2L_{2}{\chi}^{\dagger}_{1}\sigma_{2}
\tau_{2}{\chi}^{T \dagger}_{2}
({\epsilon_{(d)}}_{i}{\epsilon_{(\gamma)}}_{i}\hat{k}_{j}\hat{p}_{j} 
- {\epsilon_{(d)}}_{i}{\epsilon_{(\gamma)}}_{j}\hat{k}_{i}\hat{p}_{j}) 
\nonumber \\ && 
+ h^{1}_d\, d'_{t}2L_{2}{\chi}^{\dagger}_{1}\sigma_{i}\sigma_{2}
\tau_{3}\tau_{2}{\chi}^{T \dagger}_{2}
(\epsilon_{ijk}{\epsilon_{(d)}}_{a}
{\epsilon_{(\gamma)}}_{a}\hat{k}_{k}\hat{p}_{j}
- \epsilon_{ijk}{\epsilon_{(d)}}_{a}{\epsilon_{(\gamma)}}_{k}
\hat{k}_{a}\hat{p}_{j})\bigg] \,,
\\
iA_{PV}(c) &=& 
C d'_{t} \bigg{[} h^{1}_d {\chi}^{\dagger}_{1} 
\sigma_{i}\sigma_{2}\tau_{2}{\chi}^{T \dagger}_{2}
\epsilon_{abc}{\epsilon_{(d)}}_{c}{\epsilon_{(\gamma)}}_{a}
\hat{k}_{i}\hat{k}_{b}\mu_{V}
(m_{N}f_{1} - \gamma - ip) 
\nonumber \\
& & - h^{0t}_d {\chi}^{\dagger}_{1}
\sigma_{i}\sigma_{2}\tau_{2}{\chi}^{T \dagger}_{2}
\epsilon_{ijk}{\epsilon_{(d)}}_{a}{\epsilon_{(\gamma)}}_{k}
\hat{k}_{j}\hat{k}_{a}
\mu_{S} (m_{N}f_{1} - \gamma - ip)  
\nonumber \\
& & - h^{1}_d {\chi}^{\dagger}_{1}
\sigma_{i}\sigma_{2}\tau_{2}{\chi}^{T \dagger}_{2}
\epsilon_{ijk}{\epsilon_{(d)}}_{j}{\epsilon_{(\gamma)}}_{k}
f_{2} \bigg{]} 
\nonumber \\
&+& C d'_{s} \bigg{[} h^{1}_d (i){ \chi}^{\dagger}_{1} 
\sigma_{2}\tau_{3}\tau_{2}{\chi}^{T \dagger}_{2}
{\epsilon_{(d)}}_{i}{\epsilon_{(\gamma)}}_{i}
\mu_{S} \left\{\left(m_{N}+\frac{1}{2}\omega_{\gamma}\right)f_{1} 
- \gamma - ip\right\}  
\nonumber \\
& & + h^{0t}_d (i) {\chi}^{\dagger}_{1} \sigma_{2}\tau_{3}\tau_{2}
{\chi}^{T \dagger}_{2}{\epsilon_{(d)}}_{i}{\epsilon_{(\gamma)}}_{i}
f_{2} \bigg{]}  \,,
\\
iA_{PV}(d) &=&
C d'_{t} \bigg{[} h^{0t}_d {\chi}^{\dagger}_{1} 
\sigma_{i}\sigma_{2}\tau_{2}{\chi}^{T \dagger}_{2}
\epsilon_{abc}{\epsilon_{(d)}}_{c}{\epsilon_{(\gamma)}}_{a}
\hat{k}_{i}\hat{k}_{b}\mu_{S}
\left\{\left(m_{N}+\frac{1}{2}\omega_{\gamma}\right)f_{1}-\gamma-
ip\right\}  
\nonumber \\
& & - h^{1}_d {\chi}^{\dagger}_{1}
\sigma_{i}\sigma_{2}\tau_{2}{\chi}^{T \dagger}_{2}
\epsilon_{ijk}{\epsilon_{(d)}}_{a}{\epsilon_{(\gamma)}}_{k}
\hat{k}_{j}\hat{k}_{a}\mu_{V} 
\left\{\left(m_{N}+\frac{1}{2}\omega_{\gamma}\right)f_{1} 
- \gamma - ip\right\}  
\nonumber \\
& & - h^{1}_d {\chi}^{\dagger}_{1}
\sigma_{i}\sigma_{2}\tau_{2}{\chi}^{T \dagger}_{2}
\epsilon_{ijk}{\epsilon_{(d)}}_{j}{\epsilon_{(\gamma)}}_{k}
f_{2} \bigg{]}
\nonumber \\
&+& C d'_{s} \bigg{[} h^{0s}_d (i){ \chi}^{\dagger}_{1} 
\sigma_{2}\tau_{3}\tau_{2}{\chi}^{T \dagger}_{2}
{\epsilon_{(d)}}_{i}{\epsilon_{(\gamma)}}_{i}
\mu_{V} \left\{\left(m_{N}+\frac{1}{2}\omega_{\gamma}\right)f_{1} 
- \gamma - ip\right\}  
\nonumber \\
& & + h^{0s}_d  (i) {\chi}^{\dagger}_{1} \sigma_{2}\tau_{3}\tau_{2}
{\chi}^{T \dagger}_{2}{\epsilon_{(d)}}_{i}{\epsilon_{(\gamma)}}_{i}
f_{2}  \bigg{]}  \,,
\\
iA_{PV}(e) &=& 
Cf_{1}p[
h^{0s}_d\, d'_{s}\mu_{V}{\chi}^{\dagger}_{1}\sigma_{i}\sigma_{2}
\tau_{3}\tau_{2}{\chi}^{T \dagger}_{2}
\epsilon_{abc}{\epsilon_{(d)}}_{c}
{\epsilon_{(\gamma)}}_{a}\hat{k}_{b}\hat{p}_{i} 
\nonumber \\
& & + h^{0t}_d\, d'_{t}\mu_{S}(i){\chi}^{\dagger}_{1}\sigma_{2}
\tau_{2}{\chi}^{T \dagger}_{2}
({\epsilon_{(d)}}_{i}{\epsilon_{(\gamma)}}_{i}\hat{k}_{j}
\hat{p}_{j} - {\epsilon_{(d)}}_{i}{\epsilon_{(\gamma)}}_{j}
\hat{k}_{i}\hat{p}_{j})  
\nonumber \\
& & + h^{1}_d\, d'_{t}\mu_{S}{\chi}^{\dagger}_{1}\sigma_{i}\sigma_{2}
\tau_{3}\tau_{2}{\chi}^{T \dagger}_{2}
(\epsilon_{ijk}{\epsilon_{(d)}}_{a}{\epsilon_{(\gamma)}}_{a}
\hat{k}_{k}\hat{p}_{j}
- \epsilon_{ijk}{\epsilon_{(d)}}_{a}{\epsilon_{(\gamma)}}_{k}
\hat{k}_{a}\hat{p}_{j})]  \,,
\\
iA_{PV}(f) &=& 
C [ h^{0t}_d
(i){\chi}^{\dagger}_{1}\sigma_{2}\tau_{3}\tau_{2}
{\chi}^{T \dagger}_{2}{\epsilon_{(d)}}_{i}{\epsilon_{(\gamma)}}_{i}
- h^{1}_d{\chi}^{\dagger}_{1}\sigma_{i}\sigma_{2}\tau_{2}{\chi}^{T \dagger}_{2}
\epsilon_{ijk}{\epsilon_{(d)}}_{j}{\epsilon_{(\gamma)}}_{k}]  \,,
\\
iA_{PV}(g) &=& 
C p[h^{0t}_d\, d'_{s}
{\chi}^{\dagger}_{1}\sigma_{2}\tau_{3}\tau_{2}
{\chi}^{T \dagger}_{2}{\epsilon_{(d)}}_{i}{\epsilon_{(\gamma)}}_{i}
+ h^{1}_d\, d'_{t}(i){\chi}^{\dagger}_{1}\sigma_{i}\sigma_{2}
\tau_{2}{\chi}^{T \dagger}_{2}
\epsilon_{ijk}{\epsilon_{(d)}}_{j}{\epsilon_{(\gamma)}}_{k}] \,,
\\
iA_{PV}(h) &=& 
C\gamma[ h^{0s}_d\, d'_{s}
(i){\chi}^{\dagger}_{1}\sigma_{2}\tau_{3}\tau_{2}
{\chi}^{T \dagger}_{2}{\epsilon_{(d)}}_{i} {\epsilon_{(\gamma)}}_{i} 
- h^{1}_d\, d'_{t}{\chi}^{\dagger}_{1}\sigma_{i}\sigma_{2}
\tau_{2}{\chi}^{T \dagger}_{2}
\epsilon_{ijk}{\epsilon_{(d)}}_{j}{\epsilon_{(\gamma)}}_{k}]\,,
\end{eqnarray}
where
%
%
\begin{eqnarray}
C &=& 
\frac{1}{2}\sqrt{\frac{\gamma \rho_{d}}{1-\gamma \rho_{d}}}
\frac{1}{2\sqrt{2}\rho_d m^{5/2}_{N}} , 
\\
f_1 &=& 
\arccos \left( 
\frac{m_{N}}{\sqrt{(m_{N}+\frac{1}{2}\omega_{\gamma})^{2}-p^{2}}} 
\right)
+ i \ln \left(\frac{m_{N}+\frac{1}{2}\omega_{\gamma} + p}
{\sqrt{(m_{N}+\frac{1}{2}\omega_{\gamma})^{2}-p^{2}}} \right) , 
\\
f_2 &=& \frac{1}{\omega_{\gamma}}
[m_{N} \gamma + i(m_{N} + \frac{1}{2}\omega_{\gamma})p - 
\{(m_{N}+\frac{1}{2}\omega_{\gamma})^{2}-p^{2}\}f_{1}] , 
\\
d'_{s} &=& \frac{1}{\frac{1}{a_{0}}+ip-\frac{1}{2}r_{0}p^{2}} , 
\ \ \
d'_{t} = \frac{1}{\gamma+ip-\frac{1}{2}\rho_{d}(\gamma^{2}+p^{2})}.
\end{eqnarray}

\subsection{PV-PC interference terms}
\label{appendixB}

We have the transition rate
\begin{eqnarray}
\lefteqn{
S^{-1} \sum_{spin}^{P} |A|^{2} 
= 4(|X_{MS}|^{2}+ |Y_{MV}|^{2}-2Y_{MV} {\rm Re} X_{MS}) 
} \nonumber \\
&& + 2(|X_{MV}|^{2}+ |Y_{MS}|^{2}-2Y_{MS}{\rm Re}X_{MV}) 
+ 3(1-(\hat{k}\cdot \hat{p})^{2})(|X_{E}|^2+|Y_{E}|^{2}-2X_{E}Y_{E}) 
\nonumber \\ && 
\pm 2 \hat{n} \cdot (\hat{k} \times \hat{p})(Y_{E}-X_{E}){\rm Im}X_{MV} 
\nonumber \\ && 
\pm i (\hat{k} \cdot \hat{n}) [\tilde{f}_1-\tilde{f}_1^{*}] 
\pm i (\hat{p} \cdot \hat{k})(\hat{k} \cdot \hat{n}) 
[\tilde{f}_2-\tilde{f}_2^{*}] 
\nonumber \\ && 
\pm i (\hat{p} \cdot \hat{n}) 
[\tilde{f}_3-\tilde{f}_3^{*}] 
\pm i (\hat{p} \cdot \hat{k})(\hat{p} \cdot \hat{n}) 
[\tilde{f}_4-\tilde{f}_4^{*}]
\,.
\end{eqnarray}

PV-PC interference terms $\tilde{f}_i$ are given as
\begin{eqnarray}
\tilde{f}_1 &=& [h^{0t}_d \tilde{Z}^{pg}_{MS} - h^{0s}_d\tilde{Z}^{pg}_{MV} 
+ h^{1}_d \tilde{Z}^{pg}_{MS} ] (X^{*}_{E} - Y^{*}_{E}) \nonumber \\
& & -[2h^{0t}_d\mu_{S} \tilde{Z}^{tg} + h^{1}_d(\tilde{X}_{pp} - \tilde{Y}_{pp} 
+ 2\mu_{V}\tilde{X}_{gg} - 2\mu_{S}\tilde{Y}_{gg} 
+ 2\tilde{Z}^1_{E} + 2\mu_{V} \tilde{Z}^{t})
]X^{*}_{MV} \nonumber \\
& & +[2h^{0t}_d (\tilde{Z}^{0t}_{E} + \mu_{S}\tilde{Z}^{t}) 
+ 2h^{0s}_d (\tilde{Z}^{0s}_{E} + \mu_{V} \tilde{Z}^{sg}) 
- 2h^{1}_d (\tilde{Z}^{1}_{E} - \mu_{V}\tilde{Z}^{tg} 
- \mu_{S}\tilde{Z}^{s})]Y^{*}_{MV} \nonumber \\
& & + [h^{0t}_d(\tilde{X}_{pp} - \tilde{Y}_{pp} 
- 2\mu_{S}\tilde{X}_{gg} + 2\mu_{V}\tilde{Y}_{gg}
- 2\tilde{Z}^{0t}_{E} - 2\mu_{S}\tilde{Z}^{t}) 
- 2h^{0s}_d (\tilde{Z}^{0s}_{E} +\mu_{V}\tilde{Z}^{sg}) \nonumber \\
& & + h^{1}_d(\tilde{X}_{pp} - \tilde{Y}_{pp} 
- 2\mu_{S}\tilde{X}_{gg} + 2\mu_{V}\tilde{Y}_{gg} 
+ 2\tilde{Z}^{1}_{E} - 2\mu_{V}\tilde{Z}^{tg} - 2\mu_{S}\tilde{Z}^{s} )
]X^{*}_{MS} \nonumber \\
& & + [2h^{0t}_d \mu_{S}\tilde{Z}^{tg} 
+ 2h^{1}_d(\tilde{Z}^{1}_{E} + \mu_{V}\tilde{Z}^{t})]Y^{*}_{MS} , \\
\tilde{f}_2 &=& 
[- h^{0t}_d(\tilde{Z}^{0t}_{E} + \mu_{S}\tilde{Z}^{t} 
- \mu_{S}\tilde{Z}^{tg}) - h^{0s}_d(\tilde{Z}^{0s}_{E} 
+ \mu_{V}\tilde{Z}^{sg}) \nonumber \\ & & 
+ h^{1}_d (2\tilde{Z}^{1}_{E} + \mu_{V}\tilde{Z}^{t} 
- \mu_{S}\tilde{Z}^{s} - \mu_{V}\tilde{Z}^{tg})](X^{*}_{E} - Y^{*}_{E}) 
\nonumber \\
& & + [h^{0t}_d(\mu_{V}\tilde{X}_{pg} - \mu_{S}\tilde{Y}_{pg} ) 
- h^{1}_d(\tilde{X}_{pg} - \tilde{Y}_{pg} 
+ \mu_{S}\tilde{X}_{pg} - \mu_{V}\tilde{Y}_{pg}) ]X^{*}_{MV} \nonumber \\
& & - [h^{0t}_d \tilde{Z}^{pg}_{MS} 
- h^{1}_d \tilde{Z}^{pg}_{MS}]Y^{*}_{MV} \nonumber \\
& & + [h^{0t}_d (\tilde{X}_{pg} - \tilde{Y}_{pg} 
- \mu_{V} \tilde{X}_{pg} + \mu_{S}\tilde{Y}_{pg} 
+ \tilde{Z}^{pg}_{MS}) \nonumber \\
& & + h^{1}_d(\tilde{X}_{pg} - \tilde{Y}_{pg} 
- \mu_{V}\tilde{X}_{pg} + \mu_{S}\tilde{Y}_{pg} 
+ 2\mu_{S}\tilde{X}_{pg} - 2\mu_{V}\tilde{Y}_{pg} 
- \tilde{Z}^{pg}_{MS})]X^{*}_{MS} , \\
\tilde{f}_3 
&=& [h^{0t}_d(\tilde{Z}^{0t}_{E} + \mu_{S}\tilde{Z}^{t} 
- \mu_{S}\tilde{Z}^{tg}) + h^{0s}_d(\tilde{Z}^{0s}_{E} 
+ \mu_{V}\tilde{Z}^{sg}) \nonumber \\
& & - h^{1}_d(2\tilde{Z}^{1}_{E} - \mu_{V}\tilde{Z}^{tg} 
- \mu_{S}\tilde{Z}^{s} + \mu_{V}\tilde{Z}^{t})](X^{*}_{E} - Y^{*}_{E}) 
\nonumber \\
& & - [h^{0t}_d(\mu_{V}\tilde{X}_{pg} - \mu_{S}\tilde{Y}_{pg}) 
- 2h^{0s}_d\tilde{Z}^{pg}_{MV}  
- h^{1}_d(\tilde{X}_{pg} - \tilde{Y}_{pg} 
- \mu_{S}\tilde{X}_{pg} + \mu_{V}\tilde{Y}_{pg})]X^{*}_{MV} \nonumber \\
&& - [h^{0t}_d \tilde{Z}^{pg}_{MS} 
+ 3h^{1}_d \tilde{Z}^{pg}_{MS}]Y^{*}_{MV} \nonumber \\
&& - [h^{0t}_d (\tilde{X}_{pg} - \tilde{Y}_{pg} 
+ \mu_{V}\tilde{X}_{pg} - \mu_{S}\tilde{Y}_{pg} 
- \tilde{Z}^{pg}_{MS}) \nonumber \\
& & + h^{1}_d (\tilde{X}_{pg} - \tilde{Y}_{pg} 
+ \mu_{V}\tilde{X}_{pg} - \mu_{S}\tilde{Y}_{pg} 
+ 2\mu_{S}\tilde{X}_{pg} - 2\mu_{V}\tilde{Y}_{pg} 
- 3\tilde{Z}^{pg}_{MS})]X^{*}_{MS} \nonumber \\
&& - [2h^{0s}_d \tilde{Z}^{pg}_{MV}]Y^{*}_{MS}, \\
\tilde{f}_4 &=& 
[- h^{0t}_d \tilde{Z}^{pg}_{MS} + h^{0s}_d \tilde{Z}^{pg}_{MV} 
- h^{1}_d \tilde{Z}^{pg}_{MS}](X^{*}_{E} - Y^{*}_{E}) \nonumber \\
& &+ [ h^{1}_d (\tilde{X}_{pp} - \tilde{Y}_{pp}) ]X^{*}_{MV} \nonumber \\
&&- [h^{0t}_d(\tilde{X}_{pp} - \tilde{Y}_{pp}) 
+ h^{1}_d(\tilde{X}_{pp} - \tilde{Y}_{pp}) ]X^{*}_{MS},
\end{eqnarray}
where
\begin{eqnarray}
\tilde{X}_{pg} &=& C\frac{1}{m_{N} \omega_{\gamma}}(p \omega_{\gamma}) F^{+}, 
\ \ \ 
\tilde{Y}_{pg} = C\frac{1}{m_{N} \omega_{\gamma}}(p \omega_{\gamma}) F^{-}, 
\\
\tilde{X}_{pp} &=& C\frac{1}{m_{N} \omega_{\gamma}}(2p^{2}) F^{+}, 
\ \ \ 
\tilde{Y}_{pp} = C\frac{1}{m_{N} \omega_{\gamma}}(2p^{2}) F^{-}, 
\\
\tilde{X}_{gg} &=& C\frac{1}{m_{N} \omega_{\gamma}}
\left(\frac{{\omega_{\gamma}}^{2}}{2}\right) F^{+}, 
\ \ \ 
\tilde{Y}_{gg} = C\frac{1}{m_{N} \omega_{\gamma}}
\left(\frac{{\omega_{\gamma}}^{2}}{2}\right) F^{-},
\\
\tilde{Z}^{0s}_{E} &=& C d'_{s}[f_{2} + \gamma] \,,
\ \ \
\tilde{Z}^{0t}_{E} = Cd'_{s}
\left[f_{2} + \frac{1}{a_{0}}-\frac{1}{2}r_{0}p^{2}\right] \,,
\\
\tilde{Z}^{1}_{E} &=& -C d'_{t}
\left[2f_{2} + 2\gamma-\frac{1}{2}\rho_{d}({\gamma}^{2} + p^{2})\right] 
\,,
\\
\tilde{Z}^{sg} &=& C d'_{s}
\left[\left(m_{N} + \frac{1}{2}\omega_{\gamma}\right)f_{1}-\gamma-ip\right]\,, 
\ \ \
\tilde{Z}^{s} = C d'_{s}[m_{N}f_{1} - \gamma - ip ], 
\\
\tilde{Z}^{tg} &=& C d'_{t}
\left[\left(m_{N} + \frac{1}{2}\omega_{\gamma}\right)f_{1}-\gamma-ip \right], 
\ \ \
\tilde{Z}^{t} = C d'_{t}[m_{N}f_{1} - \gamma - ip ], 
\\
\tilde{Z}^{pg}_{MS} &=& 
C d'_{t} p  [\mu_{S} f_{1} + 2\omega_{\gamma} L_{2}], 
\ \ \
\tilde{Z}^{pg}_{MV} = C d'_{s} p [\mu_{V} f_{1} + \omega_{\gamma} L_{1}]. 
\end{eqnarray}

\subsection{$c^T_z$ and $c^T_x$}
\label{AppendixC}
\begin{eqnarray}
c^{0t}_{z} &=& - \frac{2}{\Sigma_{PC}} {\rm Im} \bigg[
\sin^2\theta(\tilde{Z}^{0t}_E + \mu_S \tilde{Z}^t - \mu_S \tilde{Z}^{tg}) (X^*_E - Y^*_E) 
\nonumber \\
& & + \{ -2\cos\theta \mu_S\tilde{Z}^{tg}
-\sin^2\theta(\mu_V \tilde{X}_{pg}-\mu_S\tilde{Y}_{pg}) \} X^*_{MV}
\nonumber \\
& & + \{ 2 \cos\theta(\tilde{Z}^{0t}_E + \mu_S\tilde{Z}^t)
-(1+\cos^2\theta)\tilde{Z}^{pg}_{MS}\} Y^*_{MV}
\nonumber\\
& & + \{ 2\cos\theta(-\mu_S\tilde{X}_{gg}+\mu_V \tilde{Y}_{gg}-\tilde{Z}^{0t}_E-\mu_S\tilde{Z}^t)
-\sin^2\theta(\tilde{X}_{pg}-\tilde{Y}_{pg})
\nonumber \\ & & 
+(1+\cos^2\theta)(-\mu_V\tilde{X}_{pg} +\mu_S\tilde{Y}_{pg}+\tilde{Z}^{pg}_{MS})\} X^*_{MS}
+ 2 \cos\theta \mu_S \tilde{Z}^{tg} Y^*_{MS}\bigg], \\
c^{0s}_z &=& - \frac{2}{\Sigma_{PC}} {\rm Im} \bigg[
\sin^2\theta (\tilde{Z}^{0s}_E +\mu_V \tilde{Z}^{sg}) (X^*_E - Y^*_E) 
+ 2\tilde{Z}^{pg}_{MV} X^*_{MV}
\nonumber \\
& & + 2\cos\theta (\tilde{Z}^{0s}_E +\mu_V \tilde{Z}^{sg}) Y^*_{MV}
- 2\cos\theta (\tilde{Z}^{0s}_E +\mu_V \tilde{Z}^{sg}) X^*_{MS}
- 2 \tilde{Z}^{pg}_{MV} Y^*_{MS}\bigg], \\
c^1_z &=& - \frac{2}{\Sigma_{PC}} {\rm Im} \bigg[
\sin^2\theta(-2\tilde{Z}^1_E+\mu_V\tilde{Z}^{tg}+\mu_S\tilde{Z}^s-\mu_V\tilde{Z}^t)(X^*_E-Y^*_E)
\nonumber \\
& & + \{-2\cos\theta(\mu_V\tilde{X}_{gg}-\mu_S\tilde{Y}_{gg}+\tilde{Z}^1_E+\mu_V\tilde{Z}^t)
\nonumber \\ & &
+\sin^2\theta(\tilde{X}_{pg}-\tilde{Y}_{pg})
- (1+\cos^2\theta)(\mu_S \tilde{X}_{pg}-\mu_V\tilde{Y}_{pg}) \}X^*_{MV}
\nonumber \\ & &
+ \{-2\cos\theta(\tilde{Z}^1_E-\mu_V\tilde{Z}^{tg}-\mu_S\tilde{Z}^s)
- (3-\cos^2\theta) \tilde{Z}^{pg}_{MS} \} Y^*_{MV}
\nonumber \\ & &
+ \{ 2 \cos\theta(-\mu_S\tilde{X}_{gg}+\mu_V\tilde{Y}_{gg}+\tilde{Z}^1_E
-\mu_V \tilde{Z}^{tg}-\mu_S\tilde{Z}^s) -\sin^2\theta (\tilde{X}_{pg}-\tilde{Y}_{pg})
\nonumber \\ & &
- (1+\cos^2\theta)(\mu_V\tilde{X}_{pg}-\mu_S\tilde{Y}_{pg})
-2\sin^2\theta(\mu_S\tilde{X}_{pg}-\mu_V\tilde{Y}_{pg})
+(3-\cos^2\theta)\tilde{Z}^{pg}_{MS}\}X^*_{MS}
\nonumber \\& &
+ 2\cos\theta(\tilde{Z}^1_E +\mu_V\tilde{Z}^{t}) Y^*_{MS}\bigg], \\
c^{0t}_x &=& \frac{2}{\Sigma_{PC}} \sin\theta {\rm Im}\bigg[
\{ \tilde{Z}^{pg}_{MS}-\cos\theta(\tilde{Z}^{0t}_E+\mu_S\tilde{Z}^t
-\mu_S\tilde{Z}^{tg})\} (X^*_E-Y^*_E)
\nonumber \\ & &
+ \{ -2\mu_S\tilde{Z}^{tg}+\cos\theta(\mu_V\tilde{X}_{pg}-\mu_S\tilde{Y}_{pg})\}X^*_{MV}
+ \{2(\tilde{Z}^{0t}_E+\mu_S\tilde{Z}^t) - \cos\theta\tilde{Z}^{pg}_{MS}\} Y^*_{MV}
\nonumber \\ & &
+ \{ (\tilde{X}_{pp}-\tilde{Y}_{pp}-2\mu_S\tilde{X}_{gg}+2\mu_V\tilde{Y}_{gg}
-2\tilde{Z}^{0t}_E-2\mu_S\tilde{Z}^t)
\nonumber \\ & &
+ \cos\theta (\tilde{X}_{pg}-\tilde{Y}_{pg}-\mu_V\tilde{X}_{pg}+\mu_S\tilde{Y}_{pg}
+\tilde{Z}^{pg}_{MS})\} X^*_{MS}
+ 2\mu_S \tilde{Z}^{tg} Y^*_{MS}
\bigg], \\
c^{0s}_x &=& \frac{2}{\Sigma_{PC}} \sin\theta {\rm Im}\bigg[
-\{ \tilde{Z}^{pg}_{MV}+\cos\theta(\tilde{Z}^{0s}_E+\mu_V\tilde{Z}^{sg})\}(X^*_E-Y^*_E)
\nonumber \\ & &
+ 2(\tilde{Z}^{0s}_E+\mu_V\tilde{Z}^{sg})Y^*_{MV} -2 (\tilde{Z}^{0s}_E+\mu_V\tilde{Z}^{sg})X^*_{MS}\bigg],\\
c^1_x &=& \frac{2}{\Sigma_{PC}} \sin\theta {\rm Im}\bigg[
\{\tilde{Z}^{pg}_{MS}+\cos\theta(2\tilde{Z}^1_E+\mu_V\tilde{Z}^t
-\mu_S\tilde{Z}^s-\mu_V\tilde{Z}^{tg}) \}(X^*_E-Y^*_E)
\nonumber \\ & &
- \{(\tilde{X}_{pp}-\tilde{Y}_{pp}+2\mu_V\tilde{X}_{gg}-2\mu_S\tilde{Y}_{gg}
+ 2\tilde{Z}^1_E+2\mu_V\tilde{Z}^t)
\nonumber \\ & &
+ \cos\theta(\tilde{X}_{pg}-\tilde{Y}_{pg}+\mu_S\tilde{X}_{pg}-\mu_V\tilde{Y}_{pg})\} X^*_{MV}
\nonumber \\ & &
+ \{ -2(\tilde{Z}^{1}_E -\mu_V \tilde{Z}^{tg}
- \mu_S\tilde{Z}^s)+\cos\theta \tilde{Z}^{pg}_{MS}\} Y^*_{MV}
\nonumber \\ & &
\{ (\tilde{X}_{pp}-\tilde{Y}_{pp}-2\mu_S\tilde{X}_{gg}+2\mu_V\tilde{Y}_{gg}
+2\tilde{Z}^1_E-2\mu_V\tilde{Z}^{tg}-2\mu_S\tilde{Z}^s)
\nonumber \\ & &
+ \cos\theta(\tilde{X}_{pg}-\tilde{Y}_{pg}-\mu_V\tilde{X}_{pg} +\mu_S\tilde{Y}_{pg}
+2\mu_S\tilde{X}_{pg}-2\mu_V\tilde{Y}_{pg}-\tilde{Z}^{pg}_{MS})\} X^*_{MS}
\nonumber \\ & &
+ 2(\tilde{Z}^1_E+\mu_V\tilde{Z}^t) Y^*_{MS} \bigg].
\end{eqnarray}


\begin{thebibliography}{99}
\bibitem{DDH}
B. Desplanques, J. F. Donoghue and B. R. Holstein, 
Ann. of Phys. {\bf 124}, 449 (1980).
\bibitem{plb01}
C. H. Hyun, T.-S. Park and D.-P. Min,
Phys. Lett. B {\bf 516}, 321 (2001).
\bibitem{schi04}
R. Schiavilla, J. Carlson and M. W. Paris,
Phys. Rev. C {\bf 70}, 044007 (2004).
\bibitem{plb03}
C. H. Hyun and B. Desplanques,
Phys. Lett. B {\bf 552}, 41 (2003).
\bibitem{prc03}
C.-P. Liu, C. H. Hyun and B. Desplanques,
Phys. Rev. C {\bf 68}, 045501 (2003).
\bibitem{epja05np}
C. H. Hyun, S. J. Lee, J. Haidenbauer and S. W. Hong,
Eur. Phys. J. A {\bf 24}, 129 (2005).
\bibitem{prc04}
C.-P. Liu, C. H. Hyun and B. Desplanques,
Phys. Rev. C {\bf 69}, 065502 (2004).
\bibitem{fuji04}
M. Fujiwara and A. I. Titov,
Phys. Rev. C {\bf 69}, 065503 (2004).
\bibitem{epja05dg}
C. H. Hyun, C.-P. Liu and B. Desplanques,
Eur. Phys. J. A {\bf 24S2}, 179 (2005).
\bibitem{prc06}
C.-P. Liu, C. H. Hyun and B. Desplanques,
Phys. Rev. C {\bf 73}, 065501 (2006).
\bibitem{prc08}
B. Desplanques, C. H. Hyun, S.-I. Ando and C.-P. Liu,
Phys. Rev. C {\bf 77}, 064002 (2008).
\bibitem{ever91}
P. D. Eversheim {\it et al.},
Phys. Lett. B {\bf 256}, 11 (1991).
\bibitem{knya84}
V. A. Knyazkov {\it et al.},
Nucl. Phys. A {\bf 417}, 209 (1984).
\bibitem{geri11}
M. T. Gericke {\it et al.},
Phys. Rev. C {\bf 83}, 015505 (2011).
\bibitem{zhu05}
S. L. Zhu, C. M. Maekawa, B. R. Holstein, 
M. J. Ransey-Musolf and U. van Kolck, 
Nucl. Phys. A {\bf 748}, 435 (2005).
\bibitem{girl08}
L. Girlanda, 
Phys. Rev. C {\bf 77}, 067001 (2008).
\bibitem{savage01}
M.~J. Savage, Nucl. Phys. A {\bf 695}, 365 (2001).
\bibitem{plb07}
C. H. Hyun, S. Ando and B. Desplanques,
Phys. Lett. B {\bf 651}, 257 (2007).
\bibitem{liu07} 
C.-P. Liu, Phys. Rev. C {\bf 75}, 065501 (2007).
\bibitem{phil09}
D. R. Phillips, M. R. Schindler and P. Springer, 
Nucl. Phys. A {\bf 822}, 1 (2009).
\bibitem{prc10} 
J.~W. Shin, S. Ando and C.~H. Hyun, 
Phys. Rev. C {\bf 81}, 055501 (2010).
\bibitem{schin10}
M. R. Schindler and R. P. Springer,
Nucl. Phys. A {\bf 846}, 51 (2010).
\bibitem{rus60}
 M.~L. Rustgi, W. Zernik, G. Breit and D.~J. Andrews,
Phys. Rev. {\bf 120}, 1881 (1960).
\bibitem{sch05} 
R.~Schiavilla, Phys. Rev. C {\bf 72}, 034001 (2005).
\bibitem{kukulin08}
V. I. Kukulin, I. T. Obukhovsky, V. N. Pomerantsev,
A. Fassler and P. Grabmayr,
Phys. Rev. C {\bf 77}, 041001(R) (2008).
\bibitem{ando11} 
S.-I. Ando, Y.-H. Song, C. H. Hyun and K.~Kubodera, 
Phys. Rev. C {\bf 83}, 064002 (2011).
\bibitem{ando05}
S. Ando and C.~H. Hyun,
Phys. Rev. C {\bf 72}, 014008 (2005).
\bibitem{ando06b} 
S. Ando, R. H. Cyburt, S.~W. Hong and C. H. Hyun,
Phys. Rev. C {\bf 74}, 025809 (2006).
\bibitem{ando08} 
S. Ando, J.~W. Shin, C.~H. Hyun, S.~W. Hong and K. Kubodera,
Phys. Lett. B {\bf 668}, 187 (2008).
\bibitem{bean01}
S. R. Beane and M. J. Savage,
Nucl. Phys. A {\bf 694}, 511 (2001).
\bibitem{alberi88}
J. Alberi {\it et al}.,
Can. J. Phys. {\bf 66}, 542 (1988).
\bibitem{prc12}
S.-I. Ando and C. H. Hyun,
Phys. Rev. C {\bf 86}, 024002 (2012).
\bibitem{liu13}
C.-P. Liu and C. H. Hyun,
in progress.
\end{thebibliography}
\end{document}